\documentclass[%
 reprint,
superscriptaddress,
 amsmath,amssymb,
 aps,
]{revtex4-2}

\usepackage{ulem}
\usepackage[dvipsnames]{xcolor}
\usepackage{graphicx}
\usepackage{dcolumn}
\usepackage{bm}

\begin{document}

\preprint{APS/123-QED}

\title{Joule spectroscopy of hybrid superconductor-semiconductor nanodevices}

\affiliation{Departamento de F\'{i}sica de la Materia Condensada, Universidad Aut\'{o}noma de Madrid, Madrid, Spain}
\affiliation{Departamento de F\'{i}sica Te\'{o}rica de la Materia Condensada, Universidad Aut\'{o}noma de Madrid, Madrid, Spain}
\affiliation{Condensed Matter Physics Center (IFIMAC), Universidad Aut\'{o}noma de Madrid, Madrid, Spain}
\affiliation{Center for Quantum Devices, Niels Bohr Institute, University of Copenhagen, Copenhagen, Denmark}

\author{A. Ibabe}
\thanks{These authors have contributed equally to this work.}
\affiliation{Departamento de F\'{i}sica de la Materia Condensada, Universidad Aut\'{o}noma de Madrid, Madrid, Spain}
\affiliation{Condensed Matter Physics Center (IFIMAC), Universidad Aut\'{o}noma de Madrid, Madrid, Spain}
\author{M. G\'{o}mez}
\thanks{These authors have contributed equally to this work.}
\affiliation{Departamento de F\'{i}sica de la Materia Condensada, Universidad Aut\'{o}noma de Madrid, Madrid, Spain}
\affiliation{Condensed Matter Physics Center (IFIMAC), Universidad Aut\'{o}noma de Madrid, Madrid, Spain}
\author{G. O. Steffensen}
\affiliation{Departamento de F\'{i}sica Te\'{o}rica de la Materia Condensada, Universidad Aut\'{o}noma de Madrid, Madrid, Spain}
\affiliation{Condensed Matter Physics Center (IFIMAC), Universidad Aut\'{o}noma de Madrid, Madrid, Spain}
\author{T. Kanne}
\affiliation{Center for Quantum Devices, Niels Bohr Institute, University of Copenhagen, Copenhagen, Denmark}
\author{J. Nyg\r{a}rd}
\affiliation{Center for Quantum Devices, Niels Bohr Institute, University of Copenhagen, Copenhagen, Denmark}
\author{A. Levy Yeyati}
\affiliation{Departamento de F\'{i}sica Te\'{o}rica de la Materia Condensada, Universidad Aut\'{o}noma de Madrid, Madrid, Spain}
\affiliation{Condensed Matter Physics Center (IFIMAC), Universidad Aut\'{o}noma de Madrid, Madrid, Spain}
\author{E. J. H. Lee}
\email{eduardo.lee@uam.es}
\affiliation{Departamento de F\'{i}sica de la Materia Condensada, Universidad Aut\'{o}noma de Madrid, Madrid, Spain}
\affiliation{Condensed Matter Physics Center (IFIMAC), Universidad Aut\'{o}noma de Madrid, Madrid, Spain}

\date{\today}

\begin{abstract}
Hybrid superconductor-semiconductor devices offer highly tunable platforms, potentially suitable for quantum technology applications, that have been intensively studied in the past decade. Here we establish that measurements of the superconductor-to-normal transition originating from Joule heating provide a powerful spectroscopical tool to characterize such hybrid devices. Concretely, we apply this technique to junctions in full-shell Al-InAs nanowires in the Little-Parks regime and obtain detailed information of each lead independently and in a single measurement, including differences in the superconducting coherence lengths of the leads, inhomogeneous covering of the epitaxial shell, and the inverse superconducting proximity effect; all-in-all constituting a unique fingerprint of each device and highlighting the large variability present in these systems. Besides the practical uses, our work also underscores the importance of heating in hybrid devices, an effect that is often overlooked.

\end{abstract}

\maketitle

The possibility to generate topological superconductivity in hybrid superconductor-semiconductor nanostructures \cite{Sau2010, Lutchyn2010, Oreg2010} has driven a strong interest towards this type of system in the past decade. Recent work has also targeted the development of novel quantum devices using the same combination of materials in the trivial regime \cite{Larsen2015, deLange2015, Tosi2019,Hays2021,Wesdorp2021}. Overall, research in the above directions has strongly benefited from remarkable developments in crystal growth and fabrication \cite{Chang2015, Krogstrup2015, Shabani2016, heedt_shadow-wall_2021}. 
By contrast, there is still a need for characterization tools that enable to efficiently probe the properties of the above materials, which is essential for understanding at depth the response of fabricated devices. In this work, we show that the Joule effect can be used as the basis for such a characterization tool for hybrid superconducting devices \cite{Choi2010, Tomi2021}. We demonstrate the potential of the technique by studying devices based on full-shell Al-InAs nanowires, also in the Little-Parks regime \cite{LittleParks1962}, and uncover clear signatures of disorder in the epitaxial shell, as well as device asymmetries resulting from the inverse superconducting proximity effect from normal metal contacts. Our results emphasize the high degree of variability present in this type of system, as well as the importance of heating effects in hybrid devices.

The Joule effect describes the heat dissipated by a resistor when  an electrical current flows, with a corresponding power equal to the product of the current and voltage in the resistor, $P = VI$. While Joule heating in superconducting devices is absent when the electrical current is carried by Cooper pairs, it reemerges when transport is mediated by quasiparticles. Interestingly, owing to the intrinsically poor thermal conductivity of superconductors at low temperatures, heating effects can be further amplified by the formation of bottlenecks for heat diffusion. As a result, the Joule effect can have a strong impact in the response of such devices. Indeed, heating has been identified as the culprit for the hysteretic $I-V$ characteristics of superconducting nanowires (NWs) \cite{Tinkham03} and overdamped $S-N-S$ Josephson junctions (where $S$ and $N$ stand for superconductor and normal metal, respectively) \cite{Courtois2008}, as well as for missing Shapiro steps in the latter \cite{DeCecco16}. In addition, it has been shown that the injection of hot electrons can significantly impact the Josephson effect in metallic \cite{Morpurgo1998} and in InAs NW-based devices \cite{roddaro_hot-electron_2011}, ultimately leading to the full suppression of the supercurrent for sufficiently high injected power.

\begin{figure}[b]
\includegraphics{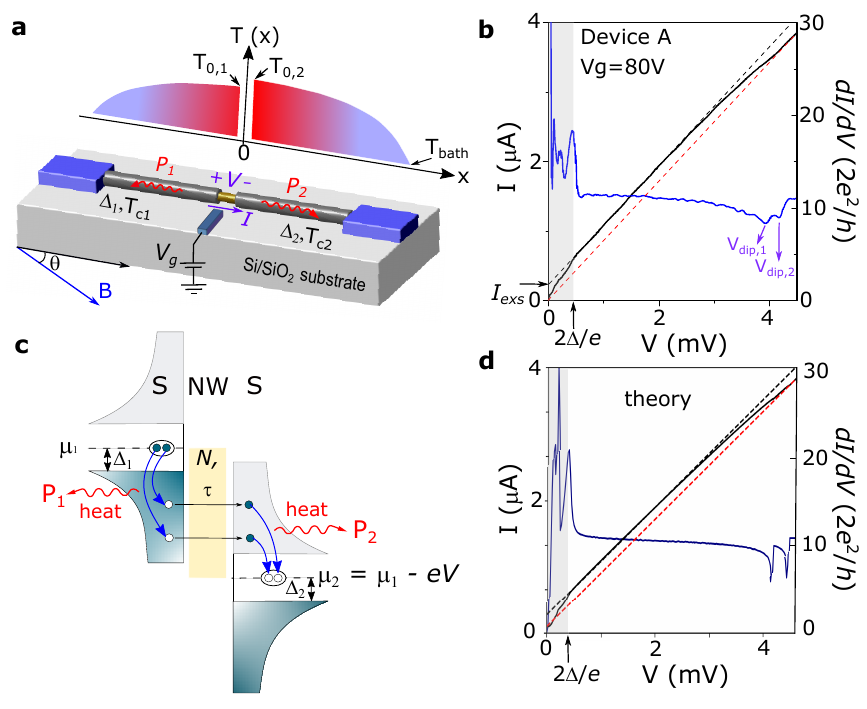}
\caption{\label{fig:principle}  $\mid$
{\textbf{Principle of Joule spectroscopy.}} \textbf{a}, Schematics of the device geometry. A Josephson junction is formed by etching a 200-nm segment of a full shell Al-InAs nanowire (NW). Voltage applied to a side gate, $V_g$, tunes the junction resistance, $R_J$. The balance between the Joule heat dissipated at the nanowire junction (equal to the product of the voltage, $V$, and current, $I$) and the cooling power from the superconducting leads 1 and 2 ($P_1$ and $P_2$) results in a temperature gradient along the device, $T(x)$. At a critical value of Joule dissipation, the temperature of the leads, $T_{0,1}$ and $T_{0,2}$, exceed the superconducting critical temperature and the leads turn normal. Each lead can display different superconducting gaps $\Delta_{1}$ and $\Delta_{2}$. An external magnetic field, $B$, is applied with an angle $\theta$ to the NW axis. $T_{bath}$ is the cryostat temperature.  \textbf{b}, $I$ (solid black line) and differential conductance, $dI/dV$ (solid blue line), as a function of $V$ measured at $V_g = 80$ V in device A. For $V < 2\Delta/e$, transport is dominated by Josephson and Andreev processes. By extrapolating the $I-V$ curve just above $V = 2\Delta/e$, an excess current of $I_{exs} \approx 200$ nA is estimated (dashed black line). Upon further increasing $V$, the Joule-mediated transition of the superconducting leads to the normal state manifest as two $dI/dV$ dips ($V_{dip, 1}$ and $V_{dip, 2}$). These transitions fully suppress $I_{exs}$ (dashed red line). \textbf{c}, The nanowire is modeled as a quasi-ballistic conductor with $N$ conduction channels with transmissions $\tau$. We assume that the energy of the quasiparticles injected in the superconductors is fully converted into heat. \textbf{d}, Keldysh-Floquet calculations of $I(V)$ and $dI/dV(V)$ using device A parameters \cite{SM}, reproducing the main features in panel \textbf{b}.
}
\end{figure}

Here, we show that instead of being merely a nuisance, Joule heating can also provide rich and independent 
information regarding each superconducting lead in hybrid superconductor-semiconductor devices in a single measurement, which can be put together to obtain a device fingerprint. To this end, we follow previous work on graphene-based Josephson junctions (JJs) \cite{Choi2010, Tomi2021} and study the Joule-driven superconductor-to-normal metal transition of the
leads in nanowire devices. Such a transition yields a clear signature in transport, namely a finite bias dip in the differential conductance, $dI/dV$, which can be used for performing spectroscopical-type measurements of the superconductivity of the leads at low temperatures. Importantly, we demonstrate that this technique, which we dub Joule spectroscopy, is able to bring to light very fine details that would otherwise be difficult to obtain only from the low-bias transport response, thus underscoring its potential for the characterization of hybrid superconducting devices. To demonstrate the technique, we focus on devices based on full-shell epitaxial Al-InAs nanowires. Specifically, we study JJs obtained by wet etching a segment of the Al shell, as schematically shown in Fig. 1a for device A (see Methods for a detailed description of the fabrication and of the different devices). For reasons that will become clearer later, we note that the leads in our JJs can display different values of superconducting critical temperature, $T_{c,i}$, and gap, $\Delta_i$, where $i$ refers to lead 1 or 2.

\begin{figure*}
\includegraphics{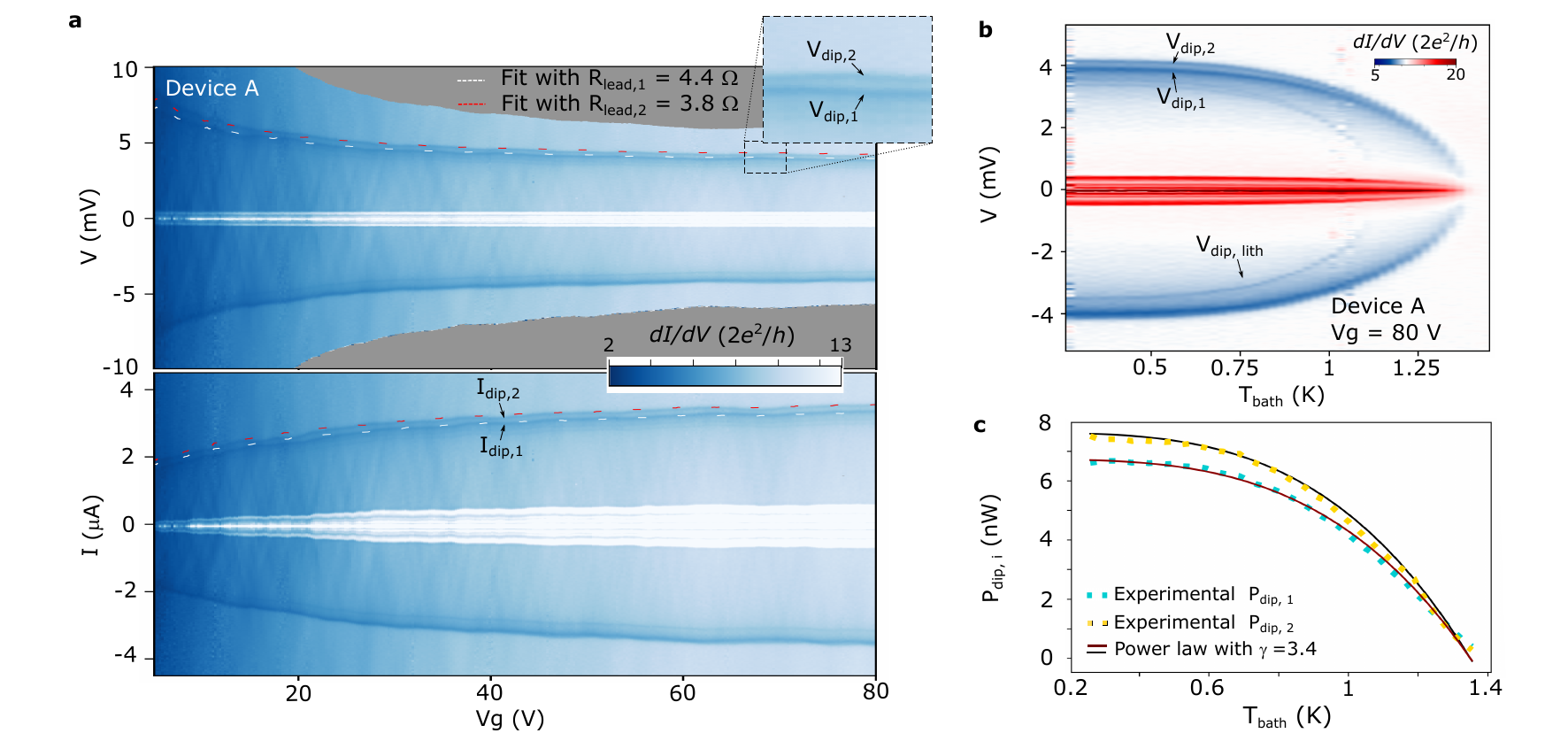} \caption{\label{fig:gate} $\mid$ \textbf{Characterization of the superconductor-to-normal metal transition of the epitaxial Al leads.} \textbf{a}, Gate voltage dependence of the $dI/dV$ for device A. The data is plotted both as a function of $V$ (top panel) and of $I$ (bottom panel). Enhanced $dI/dV$ features at low $V$ and $I$ can be attributed to Josephson and Andreev processes. Two $dI/dV$ dips, which signal the superconductor-to-normal metal transition of the leads, can be identified in each of the panels ($V_{dip, i}$ and $I_{dip, i}$). The presence of the two dips is shown in greater detail in the inset of the top panel. The white and red dashed lines are fits to Eq.~(\ref{eq:IdipVdip}) with a single free fitting parameter per lead ($R_{lead, 1}$ and $R_{lead, 2}$).
\textbf{b}, $dI/dV$ as a function of $V$ and of $T_{bath}$. A faint dip at $V_{dip, lith}$ is attributed to the Ti/Al contacts to the NW. \textbf{c},  $P_{dip,1} = V_{dip,1}I_{dip,1}/2$ (blue squares) and $P_{dip,2} = V_{dip,2}I_{dip,2}/2$ (yellow squares) as a function of $T_{bath}$. The solid lines  are fits to the power law in Eq.~(\ref{eq:powerlaw}), yielding an exponent $\gamma=3.4$.} 
\end{figure*}

\bigskip
\textbf{PRINCIPLE OF JOULE SPECTROSCOPY}
\bigskip

We start by addressing the working principle of Joule spectroscopy in greater detail. The technique relies on the balance between the Joule heat dissipated across the junction of a hybrid device and the different
cooling processes, such as electron-phonon coupling and quasiparticle heat diffusion through the leads. As both cooling processes become inefficient at low temperatures \cite{Wellstood94, Bardeen59, Knowles2012}, a heat bottleneck is established and the temperature around the junction increases (Fig.~\ref{fig:principle}a). Here, we neglect cooling by electron-phonon coupling as we estimate it to be weak \cite{SM}. We now turn to the impact of the Joule heating on the transport response of the devices. In Fig. 1b, we plot $I(V)$ and $dI/dV(V)$ traces for device A. The observed low-bias response is typical for JJs based on semiconductor nanostructures. We ascribe the $dI/dV$ peaks in this regime to a Josephson current at $V=0$ and multiple Andreev reflection (MAR) resonances at $V=2\Delta/ne$ where, for this device, $\Delta = \Delta_1 = \Delta_2 \approx 210$ $\mu$eV. Moreover, for $V \geq 2\Delta/e$, the $I-V$ curve is well described by the relation,   
\begin{equation}
I = V/R_J + I_{exs,1}(T_{0,1}) + I_{exs,2}(T_{0,2}), 
\label{eq:Iexs}
\end{equation}
where $R_J$ is the normal state junction resistance and $I_{exs,i}(T_{0,i})$ is the excess current resulting from Andreev reflections at lead $i$. Crucially, the excess current depends on the temperature of the leads at the junction, $T_{0,i}$, which can differ from each other owing to device asymmetries. For $V \lesssim 2.5$ mV, the $I_{exs,i}$ terms are approximately constant, leading to a linear $I-V$ characteristic. However, as Joule heating intensifies, deviations from this linear response follow the suppression of the excess current as $T_{0,i}$ approaches $T_{c,i}$, and $\Delta_i$ closes. At a critical voltage $V = V_{dip,i}$, the lead turns normal ($T_{0,i} = T_{c,i}$) and the excess current is fully suppressed (red dashed line in Fig. 1b), giving rise to dips in $dI/dV$  
\cite{Choi2010, Tomi2021}. We show in the following that such dips can be used for a detailed characterization of the devices.

To this end, we model the system as an $S-S$ junction with $N$ conduction channels of transmission $\tau$ connecting the two superconducting leads \cite{Goffman2017}. We further assume that injected electrons and holes equilibrate to a thermal distribution within a small distance of the junction. This is supported by the short mean-free path of the Al shell, $l \sim$ nm \cite{SM, SolePRB2020} 
, compared to the typical length of the leads, $L \sim$ $\mu$m. This equilibration results in a power, $P_i$, being deposited on either junction interface, which propagates down the Al shell by activated quasiparticles as depicted in Figs.~\ref{fig:principle}a and \ref{fig:principle}c. By solving the resulting heat diffusion equation at $T_{0,i}=T_{c,i}$, whereby we assume that the other end of the Al shell is anchored at the bath temperature of the cryostast, $T_{bath}$, 
we obtain a metallic-like Wiedemann-Franz relation for the critical power at which the dips appear \cite{SM},
\begin{equation}
P_{dip,i} = \Lambda \frac{k^2_B T^2_{c,i}}{e^2 R_{lead,i}}, \label{eq:PtoTc}
\end{equation}
where $R_{lead,i}$ is the normal resistance of the leads, and $\Lambda$ accounts for details of heat diffusion, which for the majority of experimental parameters is approximately equal to the zero-temperature BCS limit, $\Lambda \approx 2.112$ \cite{SM}.
In the high-bias limit at which the dips appear, the ohmic contribution to the current dominates $V/R_J \gg I_{exs,i}(T_{0i})$, and consequently $P_1 \approx P_2 \approx IV/2 \approx V^2/2R_J$, which implies,
\begin{equation}
V_{dip,i} = R_J I_{dip,i} = \sqrt{2\Lambda}\sqrt{\frac{R_J}{R_{lead,i}}}\frac{k_B T_{c,i}}{e},
\label{eq:IdipVdip}
\end{equation}
where $I_{dip, i}$ is the current value for the dips. Eq.~(\ref{eq:PtoTc}) and Eq.~(\ref{eq:IdipVdip}) constitute the main theoretical insights of this work and establish the basis for Joule spectroscopy. Indeed, the direct relation between $I_{dip,i}$ and $V_{dip,i}$ to  $T_{c,i}$ reveals how measurements of the dips can be used to probe the superconducting properties of the leads. To support these relations we calculate $I$ and $P_i$ self-consistently in $T_{0,i}$ by using the Floquet-Keldysh Green function technique \cite{SM}. This allows us to compare low-bias MAR structure with high-bias dip positions, and include effects of varying $\Lambda$, finite $I_{exs,i}(T_{0,i})$, pair-breaking, $\alpha$, from finite magnetic fields, and the influence of lead asymmetry on transport. Results of these calculations are shown in Fig.~{\ref{fig:principle}}d and Fig.~{\ref{fig:spec}}b with additional details given in the Supplementary Information (SI).

\begin{figure*}
\includegraphics{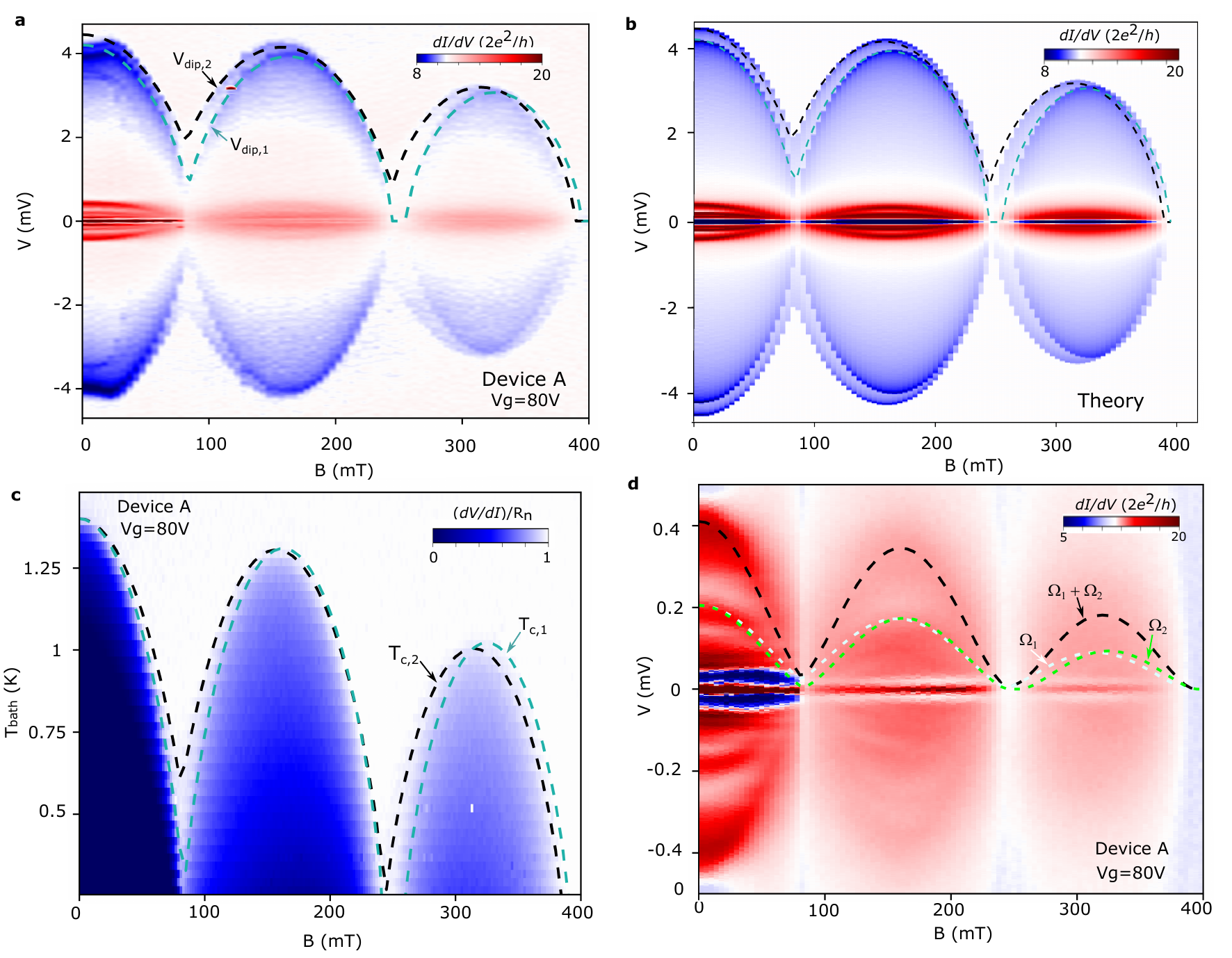} 
\caption{\label{fig:spec}  $\mid$  
\textbf{Joule effect as a spectroscopical tool.} \textbf{a}, Oscillations of $V_{dip,1}$ and $V_{dip,2}$ with applied magnetic field, 
which result from the modulation of $T_{c,i}$ by the Little Parks (LP) effect. The dashed lines are fits to the Abrikosov-Gor'kov (AG) theory, from which we conclude that the primary cause for the different LP oscillations are differences in the superconducting coherence lengths of the leads. \textbf{b}, Keldysh-Floquet calculations of the Andreev conductance at low $V$ and of the $dI/dV$ dips at high $V$ as a function of $B$ using device A parameters \cite{SM}, capturing the main experimental observations. Panels \textbf{c} and \textbf{d} demonstrate the spectroscopical potential of the technique. \textbf{c}, Zero-bias $dV/dI$ normalized by the normal state resistance of the device. The dashed lines correspond to $T_{c,i}(B)$ calculated with the AG parameters extracted by fitting the dips in panel  \textbf{a}. \textbf{d} Low-$V$ transport characterization of device A as a function of $B$. The dashed lines show the spectral gaps, $\Omega_1(B)/e$ (white) and $\Omega_2(B)/e$ (green), and their sum, $(\Omega_1(B) + \Omega_2(B))/e$ (black), obtained from $V_{dip,i}(B)$.
}
\end{figure*}

To confirm the validity of our model, we study the dependence of the dips on $R_J$, which is tuned by electrostatic gating. Following Eq.~(\ref{eq:IdipVdip}), we expect $V_{dip,i}$ ($I_{dip,i}$) to be directly (inversely) proportional to $\sqrt{R_J}$. Fig.~\ref{fig:gate}a displays $dI/dV(V)$ (top panel) and $dI/dV(I)$ (bottom panel) of device A as a function of gate voltage, $V_g$. Within the studied $V_g$ range, $R_J$ varies by a factor of $\sim$ 4. In analogy to Fig.~\ref{fig:principle}b, the high conductance regions for low $V$ ($V < 2\Delta/e$) and $I$ are due to Josephson and Andreev transport. For $V$ well above the gap, a pair of $dI/dV$ dips are detected at $V_{dip,i}$ and $I_{dip,i}$. As shown in the inset of Fig.~\ref{fig:gate}a, the two dips are better resolved for positive $V$ ($I$), reflecting a small asymmetry with respect to the sign of the bias. We fit the positions of the dips to Eq.~(\ref{eq:IdipVdip}) using $R_{lead,i}$ as a single free fitting parameter per lead/dip, as well as the experimental values for $R_J$ and $T_c = T_{c,1} = T_{c,2} = 1.35$ K. The fits, shown as white and red dashed lines in Fig.~\ref{fig:gate}a, agree remarkably well with the experimental data, thus strongly supporting our model. From these, we obtain $R_{lead, 1} = 4.4$ $\Omega$ and $R_{lead, 2} = 3.8$ $\Omega$, consistent with the normal state resistance of the Al shell ($\sim 10$ $\Omega$/$\mu$m, as measured in nominally identical NWs \cite{SM}) and lead lengths $L_{i}$ $\sim$ 0.5 $\mu$m.
The different values of $R_{lead,i}$ are attributed to slight device asymmetries, e.g., differences in $L_i$. Note that the 
good agreement of both $V_{dip,i}$ and $I_{dip,i}$ to the model demonstrates that $P_{dip,i}$ is independent of $R_{J}$, as expected from Eq.~(\ref{eq:PtoTc}) \cite{Tomi2021}.

Further information about the dips is gained by studying their dependence on $T_{bath}$. As shown in Fig.~\ref{fig:gate}b, both $V_{dip,1}$ and $V_{dip,2}$ go to zero at $T_{bath} = T_c \approx$ 1.35 K, underscoring their superconductivity-related origin. Interestingly, an additional pair of faint $dI/dV$ dips with a lower critical temperature of $T_{c,lith} \approx$ 1.1 K is observed. We conclude that these faint dips are related to the superconductivity of the lithographically-defined Al contacts shown in blue in Fig.~\ref{fig:principle}a \cite{SM}. The $T_{bath}$-dependence of the dips can also provide insights regarding the heat dissipation mechanisms of the device. As shown in Fig.~\ref{fig:gate}c, the critical power of the dips can be fitted to
\begin{equation}
\frac{P_{dip,i}(T_{bath})}{P_{dip,i}(T_{bath}=0)} = 1-(\frac{T_{bath}}{T_{c,i}})^{\gamma},
\label{eq:powerlaw}
\end{equation}
yielding $\gamma \approx 3.4$. Note that there are no additional fitting parameters to the curves and that $P_{dip,i}(T_{bath}=0)$ is calculated from the experimental $R_J$, and $R_{lead,i}$ obtained from the fits in Fig.~\ref{fig:gate}a. This is in excellent agreement with our theoretical results, from which we obtain $\gamma^{theory} \approx 3.6$ \cite{SM}. This supports our assumption that quasiparticle heat diffusion constitutes the dominant cooling mechanism in our devices.

\begin{figure*}
\includegraphics{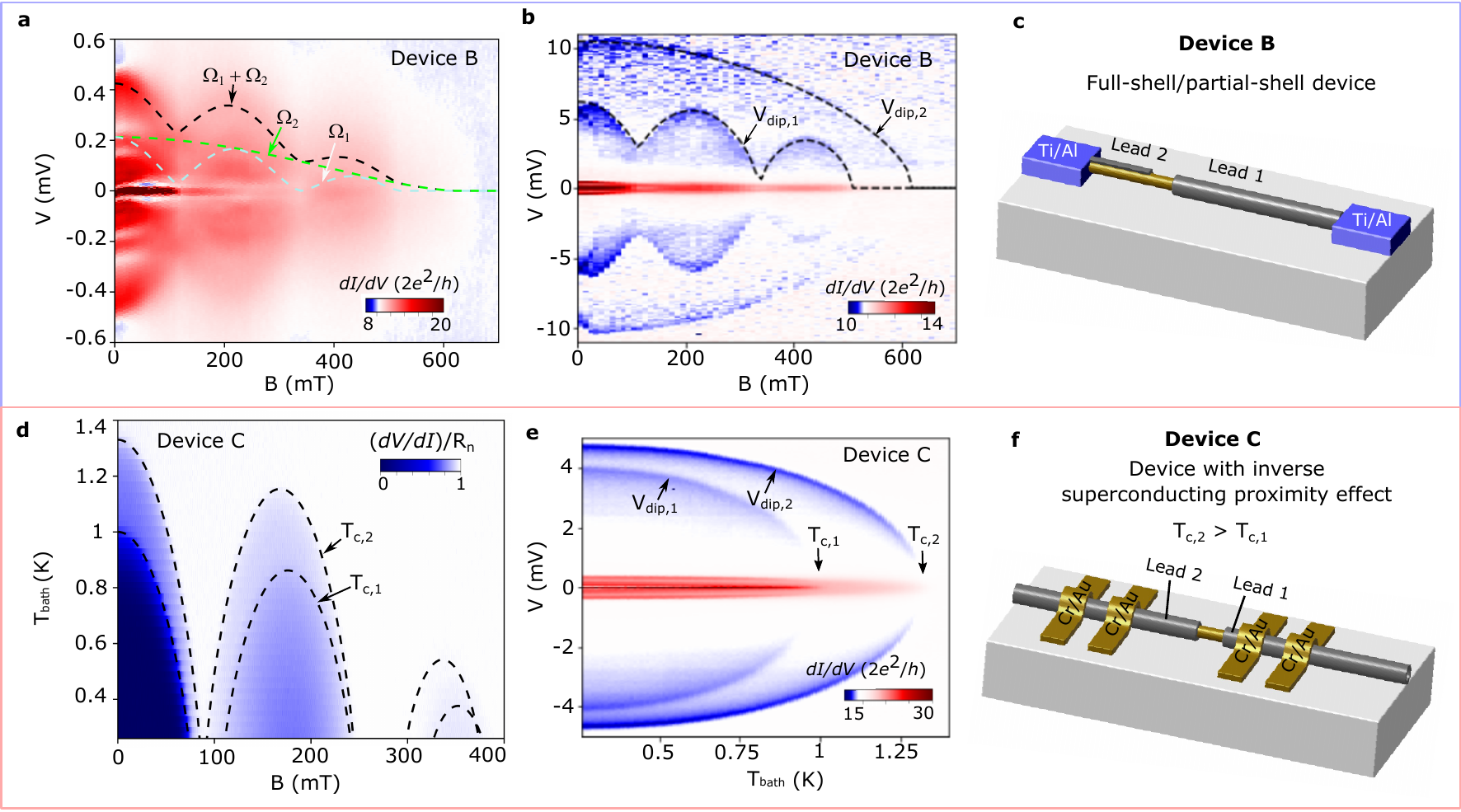}
\caption{\label{fig:example} $\mid$
\textbf{Application of Joule spectroscopy to different NW devices.} \textbf{a}, Low-bias transport characterization of device B as a function of magnetic field. Dashed lines show fittings of the spectral gaps, $\Omega_1(B)/e$ (white) and $\Omega_2(B)/e$ (green), and their sum, $(\Omega_1(B) + \Omega_2(B))/e$ (black), obtained from $V_{dip,i}(B)$. \textbf{b}, Joule spectroscopy as a function of $B$ clearly identifies that one of the superconducting leads is not doubly-connected, i.e., it behaves as a partial-shell lead. Dashed lines are fits to the AG theory. \textbf{c},  Schematics of device B, as concluded from the Joule spectroscopy characterization (not to scale). \textbf{d}, $(dV/dI)/R_n$ as a function of $T$ and $B$ for device C. The dashed lines correspond to $T_{c,i}$ obtained from $T_{c,i}(B = 0)$ and the AG fits to $V_{dip,i}(B)$ (not shown, see \cite{SM}). \textbf{e}, $T$-dependence of $V_{dip,1}$ and $V_{dip,2}$ in device C. Lead 1 displays a lower critical temperature owing to its closer proximity to the lithographic Cr/Au contacts, as depicted in the schematics in panel \textbf{f} (not to scale).}
\end{figure*}

\bigskip
\textbf{OBTAINING A DEVICE FINGERPRINT}
\bigskip

We now address the potential of Joule heating as a spectroscopical tool for hybrid superconducting devices. To accomplish this, we fix $R_J$ and study how the dips evolve as $T_{c,i}$ is tuned by an external magnetic field, $B$, approximately aligned to the NW axis (Fig.~\ref{fig:principle}a).  Fig.~\ref{fig:spec}a displays such a measurement for device A, taken at $V_g = 80$ V. Clear oscillations of $V_{dip,i}$ are observed, reflecting the modulation of $T_{c,i}$ with applied magnetic flux by the Little-Parks effect \cite{LittleParks1962,Vaitiekenas2020,vekris_asymmetric_2021, Valentini2021}. Surprisingly, the dips exhibit different Little-Parks oscillations,  suggesting that the $T_{c,i}(B)$ dependences of the two leads are not the same. To clarify this, we employ the Abrikosov-Gor'kov (AG) theory \cite{AbrikosovGorkov1961,Skalski1964Dec} to fit the experimental data (dashed lines in Fig.~\ref{fig:spec}a, see Methods for more information). 
Note that the good agreement between the dips and AG fitting is already a first indication that $V_{dip,i}$ and $T_{c,i}$ are approximately proportional, which is a consequence of
$\Lambda$ remaining nearly constant within the experimental parameter space. The discrepancies at low $B$ can be attributed to the lithographically-defined Al contacts, as we discuss in SI \cite{SM}. The AG fitting additionally reveals that the distinct dip oscillations primarily result from  differences in the superconducting coherence lengths of the leads, $\xi_{S,1} \approx 100$ nm and $\xi_{S,2} \approx 90$ nm, which owes to disorder in the epitaxial Al shell (for superconductors in the dirty limit, $\xi_S \propto \sqrt{l_e}$, where $l_e$ is the mean free path) \cite{SolePRB2020, SM}. The main features of the experimental data are well captured by the results of our Floquet-Keldysh calculations using parameters obtained from the AG fitting (Fig.~\ref{fig:spec}b). 

Further support for Joule spectroscopy is gained by verifying that $V_{dip,i}$ and $T_{c,i}$ remain proportional as a function of $B$. To this end, we measure the differential resistance, $dV/dI$, of the device at $V=0$, as shown in Fig.~\ref{fig:spec}c. Regions in which $dV/dI < R_n$, where $R_n$ is the normal state resistance, indicate that at least one of the leads is superconducting, whereupon the device conductance is enhanced either by Josephson or Andreev processes. The dashed lines correspond to the expected values of $T_{c,i}(B)$ from AG theory, which were calculated from the experimental zero-field critical temperature ($T_c = 1.35$ K) and parameters obtained from AG fitting in Fig.~\ref{fig:spec}a. A very good agreement with the experimental data is observed, also allowing to identify regions in which only one of the leads is superconducting (i.e., between the dashed lines, where $dV/dI$ takes values slightly lower than $R_n$).
This demonstrates that the linear relation between $V_{dip,i}$ and $T_{c,i}$ is preserved for experimentally-relevant conditions, as required by the technique. We also stress that while the differences in $\xi_{S,i}$ are barely visible in Fig.~\ref{fig:spec}c, they can be detected in a significantly clearer (and faster) manner using Joule spectroscopy. Overall, the above observations demonstrate the ability of the technique in obtaining a device fingerprint. We emphasize that such detailed information of the superconducting leads separately  
is not directly accessible from the low-bias transport response, which we discuss below.

We now show that the information gained from Joule spectroscopy 
provides a consistent description of the low-bias device response with respect to the experimental data (Fig.~\ref{fig:spec}d). For this comparison, we focus on MAR resonances of orders $n =1$ and 2 which, for $B = 0$, are centered at $V = (\Delta_1+\Delta_2)/e$, and $V = \Delta_1/e$ and $V = \Delta_2/e$, respectively ($\Delta_i$ are obtained from the experimental $T_{c,i}$ using the BCS relation $\Delta \approx 1.76k_{B}T_{c}$ valid at zero field). 
Owing to depairing effects, the MAR resonances cease to depend linearly on $\Delta_i$ and $T_{c,i}$ at finite $B$. Instead, the position of MAR peaks is better captured by scalings with the spectral gap, $\Omega_i(B) = \Delta_i(B = 0)(T_{c,i}(B)/T_{c,i}(B = 0))^{5/2}$, as concluded from our numerical simulations \cite{SM}. In Fig.~\ref{fig:spec}d, we plot $(\Omega_1+\Omega_2)/e$ (black), $\Omega_1/e$ (white), and $\Omega_2/e$ (green) as dashed lines, which were calculated using $T_{c,i}(B)$ extracted from the dips in Fig. \ref{fig:spec}a. Curiously, the visibility of MAR features reduces with increasing Little-Parks lobe, which makes it more difficult to compare the experimental data with the spectral gaps for $B \gtrsim$ 100 mT. Regardless, a reasonable agreement with the data is observed (more clearly seen in the zeroth lobe), even though our experiment is not able to resolve the splitting between the $\Omega_1/e$ and $\Omega_2/e$ peaks (see also Extended Data Fig. 1). 

\bigskip
\textbf{DEMONSTRATION OF LARGE DEVICE VARIABILITY}
\bigskip

Applying Joule spectroscopy to a number of different samples underscores that each device is unique. We present below two additional examples of devices based on nominally identical NWs. We start by device B, which has the same geometry as device A with the exception that the lengths of the epitaxial Al leads are made purposefully asymmetric ($L_{1(2)} \approx 0.5(0.7) \mu$m). The low-bias transport response shown in Fig.~\ref{fig:example}a is similar to that of device A, although the MAR oscillations with $B$ are not as clearly discernible. Despite the similarities, Joule spectroscopy reveals that this device is in fact quite different. It demonstrates that one of the Al leads is not doubly-connected, as concluded from the fact that only one of the dips displays the Little-Parks effect  (Fig.~\ref{fig:example}b). Such a behavior can be linked to a discontinuity in the Al shell formed either during growth or the wet etching of the shell. Note that the different values of $V_{dip,i}$ are due to differences in $R_{lead,i}$, which scale with the lead length. In analogy to device A, we compare the information gained from the dips (shown as dashed lines in Fig.~\ref{fig:example}a) with the low-bias data. We obtain a reasonable correspondence with the experimental data, including the splitting between the $\Omega_1/e$ and $\Omega_2/e$ lines, which is particularly visible in the zeroth lobe. 

In our last example, we study a device with a 4-terminal geometry and with normal (Cr/Au) electrical contacts to the Al-InAs NW (device C). $L_i$ in this device is also asymmetric (here, taken as the distance from the junction to the voltage probes). Fig.~\ref{fig:example}d displays the zero-bias $dV/dI$ of the device as a function of $T$ and $B$. At $B=0$, it is easy to identify that $dV/dI$ increases more abruptly at two given temperatures. Joule spectroscopy taken as a function of $T$ and at $B=0$ (Fig.~\ref{fig:example}e) reveals that the two superconducting leads display different critical temperatures, $T_{c, 1} \approx 1 K$ and $T_{c, 2} \approx 1.33 K$. This behavior owes to the inverse superconducting proximity, which scales inversely with the distance to the Cr/Au contacts. In analogy to device A, we fit $V_{dip,i}(B)$ with AG theory (Extended Data Fig.~1), and use the same fitting parameters to obtain $T_{c,i}(B)$, which are plotted as dashed lines in Fig.~\ref{fig:example}d. As in the previous examples, a very good agreement is obtained with the experimental data. 

\bigskip
\textbf{CONCLUSION}
\bigskip

To conclude, we have demonstrated that the Joule effect can be fostered to provide a quick and detailed fingerprint of hybrid superconductor-semiconductor devices. By studying nominally-identical Al-InAs nanowires, we observe 
that intrinsic disorder in the epitaxial shell, and extrinsic factors, such as the inverse superconducting proximity effect, inevitably contribute to making each device unique. Concretely, this results in asymmetries in the superconducting leads that often remain undetected owing to the difficulty to obtain separate information from the individual leads in low-bias measurements. We have shown that these asymmetries can be substantial, directly impacting the device response, and that they can be further amplified with external magnetic fields, a regime which has been largely explored in the past decade in the context of topological superconductivity \cite{prada_andreev_2020}. Joule spectroscopy thus constitutes a powerful, complimentary tool to low-bias transport.
Clearly, the technique is not restricted to the material platform investigated here, and will also be of use for the characterization of novel materials \cite{kanne_epitaxial_2021, Pendharkar_Sn_2021, Jung_AdvMater_growth21}. Our work also points out the importance of heating in hybrid superconducting devices. Indeed, owing to the poor thermal conductivity of superconductors, the device temperature can be considerable even at voltages way below the superconductor-to-normal metal transitions discussed here, and possibly also in microwave experiments which are currently carried out in these devices \cite{Tosi2019,Hays2021,Wesdorp2021}. To the best of our knowledge, such heating effects have not been typically taken into account in this type of systems. Further work is needed to clarify its possible consequences in device response. 
 
\bigskip

\textbf{METHODS}

\textbf{Sample fabrication and measured samples}: The devices studied in this work are based on InAs nanowires (nominal diameter, $d = 135$ nm) fully covered by an epitaxial Al shell (nominal thickness, $t = 20$ nm). The nanowires are deterministically
transferred from the growth chip to Si/SiO$_2$ (300 nm) substrates using a micro-manipulator. E-beam lithography (EBL) is then used to define a window for wet etching an approx. 200 nm-long segment of the Al shell. A 30 s descumming by oxygen plasma at 200 W is performed before
immersing the sample in the AZ326 MIF developer (containing 2.38\% tetramethylammoniumhydroxide, TMAH) for 65 s at room temperature. Electrical contacts and side gates are subsequently fabricated by standard EBL techniques, followed by ion milling to remove the oxide of the Al shell, and metallization by e-beam evaporation at pressures of $\sim 10^{-8}$ mbar. Here, we have explored devices with two different types of electrical contacts, namely superconducting Ti (2.5 nm)/Al (240 nm) or normal Cr (2.5 nm)/Au (80 nm), the latter of which were deposited by angle evaporation to ensure the continuity of the metallic films. 

Overall, we have measured a total of 18 devices from 6 different samples. The main features discussed in this work have been observed in all of the devices. We focus our discussion in the main text to data corresponding to three devices from three different samples. Device A was fabricated with superconducting Ti/Al contacts and a side gate approximately 100 nm away from the junction. The nominal lengths of its epitaxial superconducting leads were $L_1 = 0.42$ $\mu$m, $L_2 =0.45 $ $\mu$m. Device B also had superconducting Ti/Al contacts, but the charge carrier density was tuned by a global back gate (here, the degenerately-doped Si substrate, which is covered by a 300 nm-thick SiO$_2$ layer). The lengths of the epitaxial superconducting leads were made purposefully asymmetric (nominal lengths $L_1 = 0.5 $ $\mu$m, $L_2 = 0.7 $ $\mu$m) to further confirm the impact of $R_{lead,i}$ on $V_{dip,i}$. Finally, device C had a four-terminal geometry with normal Cr/Au contacts and a global back gate. The lengths of the epitaxial leads (in this case, the distance from the junction to the voltage probes) were nominally $L_1 = 0.3 $ $\mu$m, $L_2 = 0.6$ $\mu$m. 

\textbf{Measurements}: Our experiments were carried out using two different cryogenic systems: a $^3$He insert with a base temperature of 250 mK, employed in the measurements of devices A and C, and a dilution refrigerator with a base temperature of 10 mK, which was used in the measurements of device B. 

We have performed both voltage-bias (devices A and B) and current-bias (devices A and C) transport measurements using standard lock-in techniques.  Typically, for a given device, we have taken different measurements both at "low-bias" and "high-bias". The former refers to limiting $V$ and $I$ to focus on the Josephson and Andreev transport that occurs for $V \leq 2\Delta/e$. By contrast, the latter corresponds to biasing the device enough to reach the regime whereby Joule effects become significant. We have employed different levels of lock-in excitation for the "low-bias" and "high-bias" measurements. Respectively, the lock-in excitations were: $dV = 5-25$ $\mu$V and $dV = 100-200$ $\mu$V for voltage-bias measurements (note: the $dV$ values listed above are nominal, i.e., without subtracting the voltage drop on the cryogenic filters), and $dI = 2.5$ nA and $dI = 20$ nA for current-bias measurements. 

\textbf{Data processing}: The voltage drop on the total series resistance of two-terminal devices (devices A and B), which are primarily due to cryogenic filters (2.5 k$\Omega$ per experimental line), have been subtracted for plotting the data shown in Figs. 1-3 and Fig. 4a,b. 

\textbf{Data analysis}:
Following previous work on full-shell Al-InAs nanowires \cite{vekris_asymmetric_2021, SolePRB2020}, we employ a hollow cylinder model for the Al shell, assumed to be in the dirty limit, which is justified by the fact that the electron gas in Al-InAs hybrids accumulates at the metal-superconductor interface. In this geometry the application of a parallel magnetic field leads to a oscillating pair-breaking parameter \cite{Shah2007},
\begin{equation}
\alpha_{\|}=\frac{4 \xi_{S}^{2} T_{c}(0)}{A}\left[\left(n-\frac{\Phi_{\|}}{\Phi_{0}}\right)^{2}+\frac{t_{S}^{2}}{d^{2}}\left(\frac{\Phi_{\|}^{2}}{\Phi_{0}^{2}}+\frac{n^{2}}{3}\right)\right],
\end{equation}
with $n$ denoting the fluxoid quantum number, $A$ the cross-sectional area of the wire, $t_S$ the thickness of the Al shell, and $\Phi_\| = B_| A$ the applied flux. For a perpendicular field a monotone increase of pair-breaking is observed (see Extended Data Fig.~\ref{fig:PerpFields}), which we fit to the formulae of a solid wire assuming $d\lesssim\xi_S$ with $d$ denoting diameter \cite{Shah2007, vekris_asymmetric_2021, Vaitiekenas2020},
\begin{equation}
\alpha_{\perp}=\frac{4 \xi_{S}^{2} T_{c}(0)\lambda}{A} \frac{\Phi_{\perp}^{2}}{\Phi_{0}^{2}}, \label{eq:alpPerp}
\end{equation}
with $\Phi_{\perp} = B_{\perp}A$ and $\lambda$ being a fitting parameter \cite{vekris_asymmetric_2021}. In our analysis of parallel fields we include a small angle, $\theta$, between the external field and the nanowire axis, which is typically present in the experimental setup (see Fig. \ref{fig:principle}a). This angle is treated as a fitting parameter and can be distinct between lead 1 and 2 due to possible curvature of the NW. Consequently, the full pair-breaking is given by $\alpha(B) = \alpha_\|(B) + \alpha_\perp(B)$ with $B_| = B \cos\theta$ and $B_\perp = B\sin\theta$ from which we can extract the critical temperature, $T_c(\alpha)$, using AG theory,
\begin{equation}
\label{eqn:AGE}
\ln\left(\frac{T_{c}({\alpha})}{T_{c}(0)}\right) = \Psi\left(\frac{1}{2}\right)
- \Psi\left(\frac{1}{2} + \frac{\alpha}{2\pi k_{B}T_{c}(\alpha)} \right),
\end{equation}
where $\Psi$ is the digamma function. From the proportionality, $T_{c}(B)/T_{c}(0) \approx V_{dip}(B)/V_{dip}(0)$, we obtain good fits for all devices and leads assuming $t_S \approx 15$~nm \cite{SM}, close to the nominal thickness of $20$~nm from the crystal growth. This discrepancy is attributed to uncertainties in the Al deposition thickness during growth, and to the formation of an oxide layer present on all shells. 
From these fits we obtain the coherence lengths, $\xi_{S,i}$, and find distinct values for lead 1 and 2 in all devices. We note that the obtained $\xi_{S,i}$ values are in good agreement with values estimated from the mean-free path of the Al shell. From LP periodicity we extract wire diameter and find $d_A$, $d_C \approx 125$~nm and $d_B\approx 105$~nm with $A$, $B$ and $C$ indicating device. For these values $d_i \gtrsim \xi_{S,i}$, possibly leading to slight modifications of eq.~(\ref{eq:alpPerp}) which are accounted for by the fitting parameter $\lambda$. The discrepancy between the estimated values for devices $A$ and $C$ with respect to the nominal diameter are attributed to the diameter distribution obtained in the employed growth conditions. The thinner wire in device $B$, on the other hand, could result from special growth conditions (i.e., by sharing some of the substrate adatom collection area with a spurious extra wire). Further details and tables of device parameters can be found in the Supplementary Information \cite{SM}.

For finite magnetic fields, the linear BCS relation between $T_{c}(B)$ and $\Delta(B)$ is no longer valid. Our theoretical simulations indicate that in this limit, the MAR features follow the spectral gap, $\Omega(B) \approx \Delta_0(T_{c}(B)/T_{c}(0))^{5/2}$ \cite{SM}. This relation is used to fit low-bias MAR signatures from high-bias measurements of $V_{dip}$.

 \bibliography{bibliography}
 
\textbf{Author contributions}

A.I. fabricated the device, A.I., M.G., and E.J.H.L. performed the measurements and analyzed the experimental data. G.O.S. and A.L.Y. developed the theory. G.O.S. performed the theoretical calculations. T.K. and J.N. developed the nanowires. All authors discussed the results. A.I., M.G., G.O.S., A.L.Y., and E.J.H.L, wrote the manuscript with input from all authors. E.J.H.L. proposed and guided the experiment.

\begin{acknowledgments}
We wish to thank Marcelo Goffman, Hughes Pothier and Cristian Urbina for useful comments. 
We acknowledge funding by EU through the European Research Council
(ERC) Starting Grant agreement 716559 (TOPOQDot), the FET-Open contract AndQC, by the Danish National Research Foundation, Innovation Fund Denmark, the Carlsberg Foundation, and by the Spanish AEI through Grant No.~PID2020-117671GB-I00 and through the ``Mar\'{\i}a de Maeztu'' Programme for Units of Excellence in R\&D (CEX2018-000805-M) and the ''Ram\'{o}n y Cajal'' programme grant RYC-2015-17973.
\end{acknowledgments} 

\newpage
\begin{figure*}
    \centering
    \includegraphics[width=16cm]{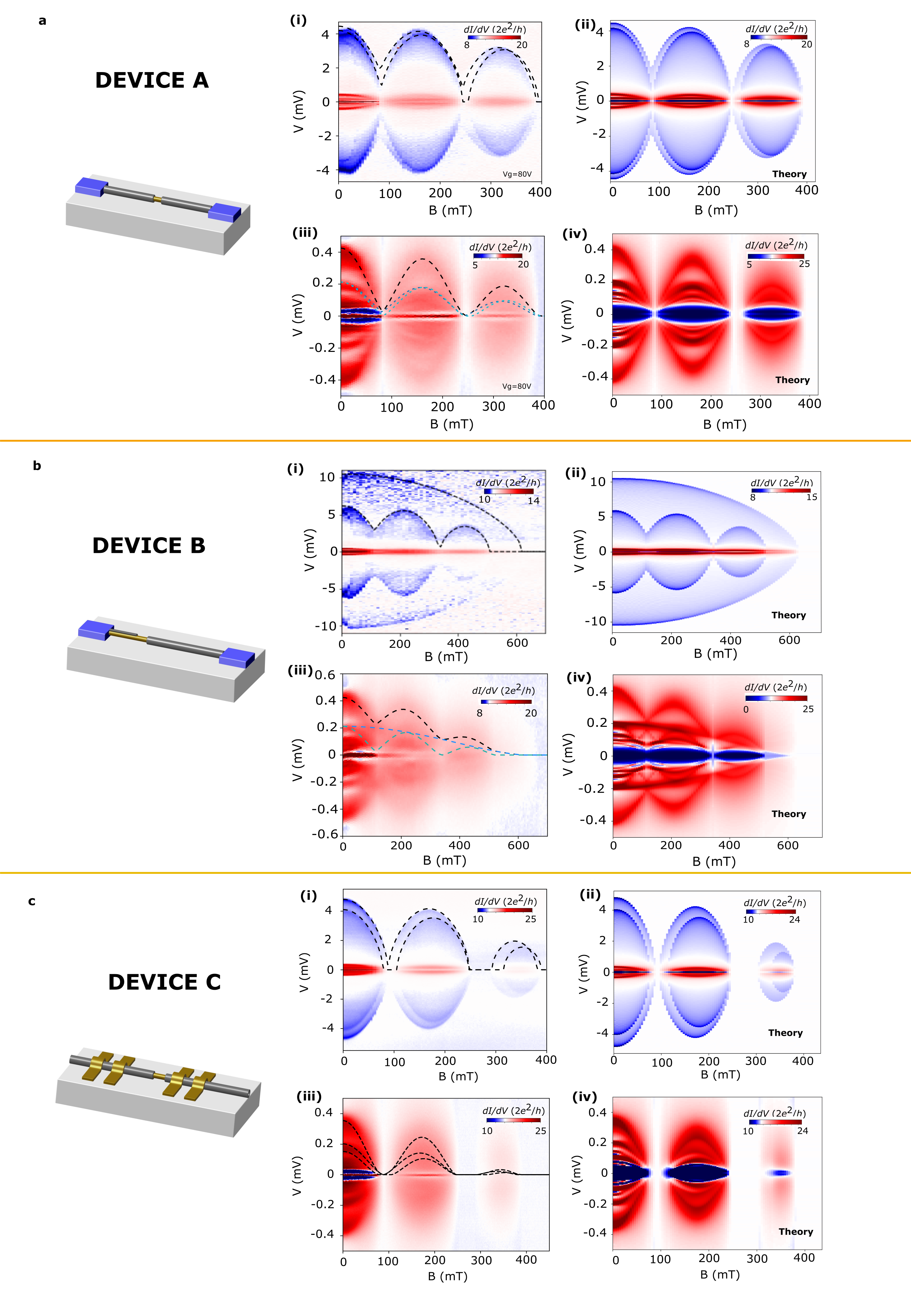}
    \caption{\textbf{Extended Data Figure 1 $\mid$ Joule spectroscopy characterization of devices A, B and C}. For each device we plot: (i) the Joule spectrum of the leads and (ii) $dI/dV$ at low-$V$ as a function of $B$, and Floquet-Keldysh calculations of the (iii) high-$V$ and (iv) low-$V$ transport response.}
    \label{fig:FullDataABC}
\end{figure*}

\newpage

\begin{figure*}
    \centering
    \includegraphics[width=16cm]{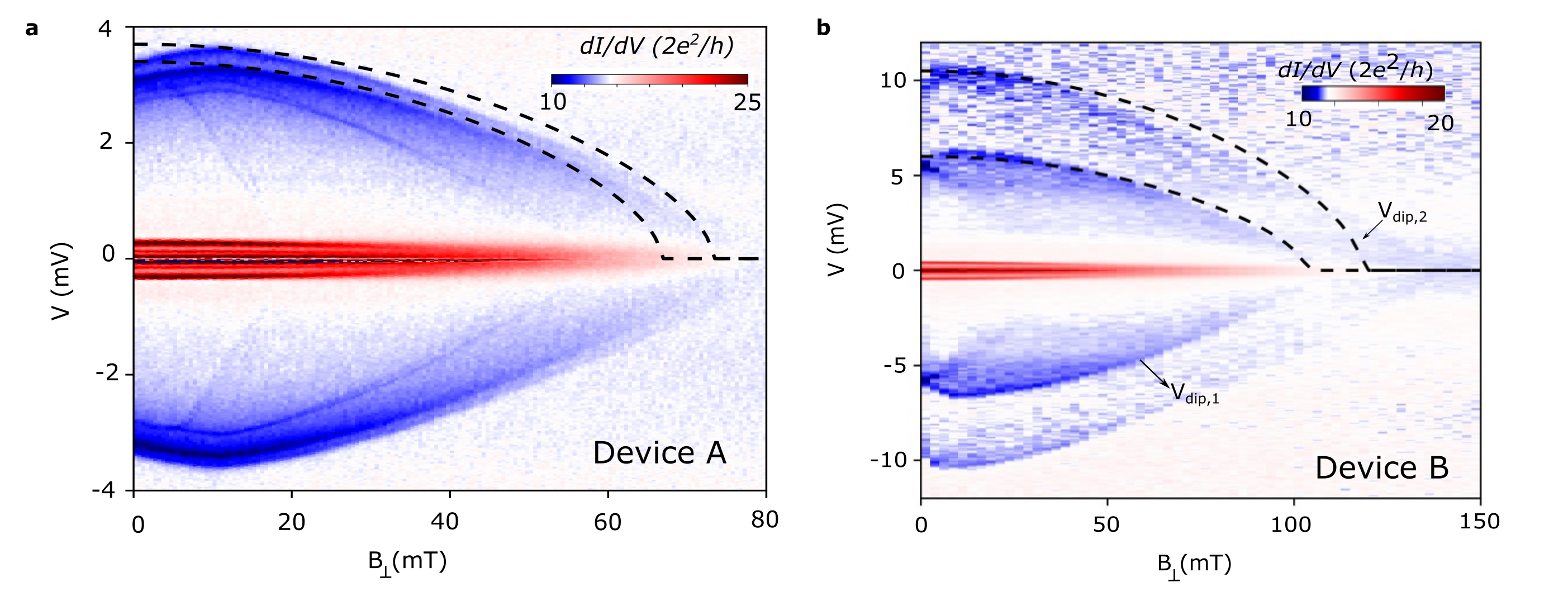}
    \caption{\textbf{Extended Data Figure 2 $\mid$ Perpendicular magnetic field dependences.}
    $dI/dV(V)$ as a function of perpendicular magnetic field, $B_{\perp}$ for devices A (panel \textbf{a}) and B (panel \textbf{b}). Dashed lines show predictions of the AG theory using the same parameters obtained from fitting the data with the nearly parallel magnetic field, $B$.
}.
    \label{fig:PerpFields}
\end{figure*}

\clearpage

\begin{figure*}
    \centering
    \includegraphics[width=16cm]{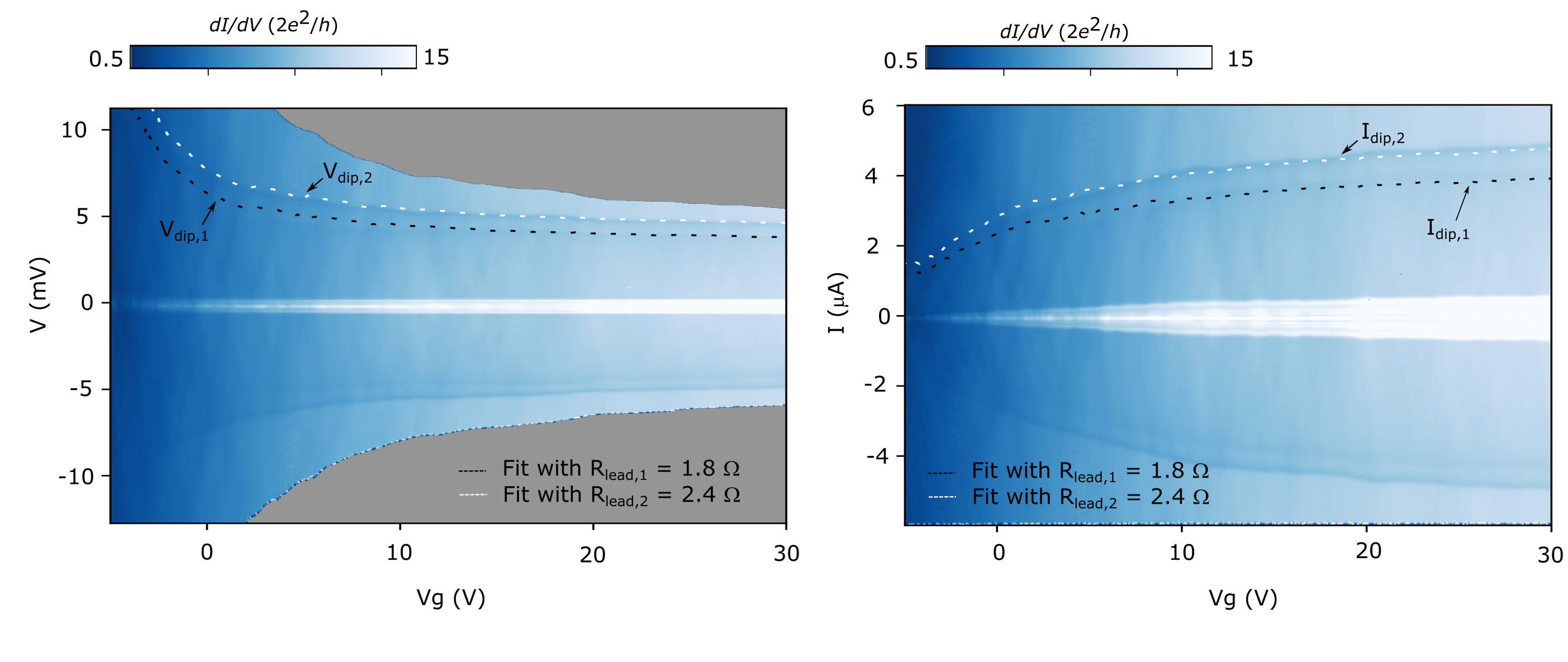}
    \caption{\textbf{Extended Data Figure 3 $\mid$ Gate dependence of the dip features in device C}. $dI/dV(V)$ (left panel) and $dI/dV(I)$ (right panel) measured in a 4-terminal configuration as a function of the gate voltage. Notice that no post-processing of the data is required to obtain the real voltage drop across the device (see Methods), as the measurement is not affected by series resistances in the experimental setup (e.g., the resistance of the cryogenic filters). The dashed lines are fits to Eq.~(3) in the main text with a single free fitting parameter per dip, $R_{lead,1}$ and $R_{lead,2}$. An excellent agreement is obtained between the experimental data and the fits, from which we obtain $R_{lead,1} \approx 1.8$ $\Omega$ and $R_{lead,2} \approx 2.4$ $\Omega$. Note that for this analysis, we have used the two different superconducting critical temperatures of the leads, namely $T_{c,1} = 0.98$ K and $T_{c,2} = 1.31$ K, which result from the inverse superconducting proximity effect.}

    \label{fig:Rn deviceC}
\end{figure*}

\end{document}


\title{\texorpdfstring{Supplementary Information \\ Joule spectroscopy of hybrid superconductor-semiconductor nanodevices}{Supplementary Information}}

\affiliation{Departamento de F\'{i}sica de la Materia Condensada, Universidad Aut\'{o}noma de Madrid, Madrid, Spain}
\affiliation{Departamento de F\'{i}sica Te\'{o}rica de la Materia Condensada, Universidad Aut\'{o}noma de Madrid, Madrid, Spain}
\affiliation{Condensed Matter Physics Center (IFIMAC), Universidad Aut\'{o}noma de Madrid, Madrid, Spain}
\affiliation{Center for Quantum Devices, Niels Bohr Institute, University of Copenhagen, Copenhagen, Denmark}

\author{A. Ibabe}
\thanks{These authors have contributed equally to this work.}
\affiliation{Departamento de F\'{i}sica de la Materia Condensada, Universidad Aut\'{o}noma de Madrid, Madrid, Spain}
\affiliation{Condensed Matter Physics Center (IFIMAC), Universidad Aut\'{o}noma de Madrid, Madrid, Spain}
\author{M. G\'{o}mez}
\thanks{These authors have contributed equally to this work.}
\affiliation{Departamento de F\'{i}sica de la Materia Condensada, Universidad Aut\'{o}noma de Madrid, Madrid, Spain}
\affiliation{Condensed Matter Physics Center (IFIMAC), Universidad Aut\'{o}noma de Madrid, Madrid, Spain}
\author{G. O. Steffensen}
\affiliation{Departamento de F\'{i}sica Te\'{o}rica de la Materia Condensada, Universidad Aut\'{o}noma de Madrid, Madrid, Spain}
\affiliation{Condensed Matter Physics Center (IFIMAC), Universidad Aut\'{o}noma de Madrid, Madrid, Spain}
\author{T. Kanne}
\affiliation{Center for Quantum Devices, Niels Bohr Institute, University of Copenhagen, Copenhagen, Denmark}
\author{J. Nyg\r{a}rd}
\affiliation{Center for Quantum Devices, Niels Bohr Institute, University of Copenhagen, Copenhagen, Denmark}
\author{A. Levy Yeyati}
\affiliation{Departamento de F\'{i}sica Te\'{o}rica de la Materia Condensada, Universidad Aut\'{o}noma de Madrid, Madrid, Spain}
\affiliation{Condensed Matter Physics Center (IFIMAC), Universidad Aut\'{o}noma de Madrid, Madrid, Spain}
\author{E. J. H. Lee}
\email{eduardo.lee@uam.es}
\affiliation{Departamento de F\'{i}sica de la Materia Condensada, Universidad Aut\'{o}noma de Madrid, Madrid, Spain}
\affiliation{Condensed Matter Physics Center (IFIMAC), Universidad Aut\'{o}noma de Madrid, Madrid, Spain}

\maketitle

\tableofcontents

\newpage

\section{Supplementary experimental data}

\subsection{Properties of the epitaxial Al shell}

We present here a characterization of the epitaxial Al shell of nanowires from the same batch as that used for devices A, B and C. We have fabricated devices with a 4-terminal geometry and with angle-evaporated Cr(2.5 nm)/Au(80 nm) contacts, similar to device C. In this case, however, the Al shell was not etched. Current-biased measurements were taken at low temperatures and with an external magnetic field, $B$. Such a characterization was aimed at estimating relevant parameters of the Al shell, such as the normal state resistance, $R_n$, the superconducting coherence length, $\xi_{S}$, and the critical current, $I_c^{shell}$, to compare with the results obtained from our analyses of the dips in the main text.  

\begin{figure}[H]
    \centering
    \includegraphics[width=16cm]{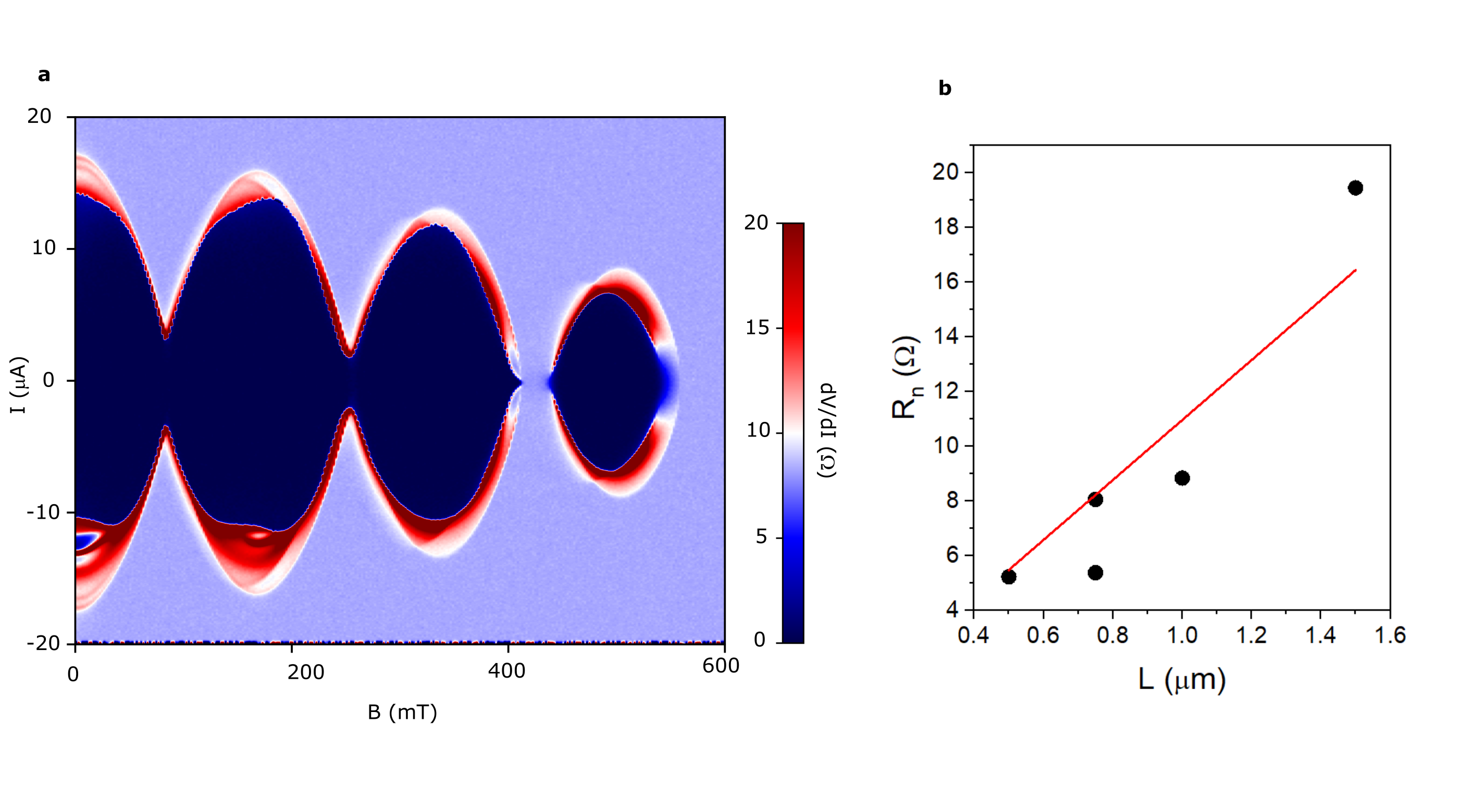}
    \caption{\textbf{Characterization of the epitaxial Al shell}. \textbf{a}, $dV/dI(I)$ measurement taken as a function of $B$. \textbf{b}, Length dependence of the normal state resistance of the Al shell. $L$ corresponds to the distance between the voltage probes of the device.}
    \label{fig:Al_shell}
\end{figure}

Fig.~\ref{fig:Al_shell} displays a typical $dV/dI (I, B)$ measurement, where $I$ is the current bias. Note that the measurements were taken by sweeping $I$ from negative to positive values and, as such, features in the negative/retrapping branch may be affected by heating effects. We will not discuss this in further detail, as it is outside of the scope of this work. By measuring a total of 5 devices, we have observed a distribution of critical current,  $I_c^{shell} \approx 10-25 \mu$A (taken at positive $I$). These values are at least 2-3 times larger than the highest values measured for $I_{dip}$, reinforcing that the reported dips are not related to the critical current of the shell. 

Concerning the normal state resistance of the shell, we define $R_n = dV/dI(I > I_c^{shell})$. In Fig.~\ref{fig:Al_shell}, we plot $R_n$ as a function of the distance between the voltage probes, $L$. By applying a linear fit to the datapoints, we estimate $R_n/L \approx 11$ $\Omega/µ$m. As mentioned in the main text, the $R_{lead}$ values obtained by fitting the dips agree very well with this estimate.  

We now evaluate the superconducting coherence length of the epitaxial shell. We estimated $\xi_{\mathrm{S}}$ from $R_n$ by applying the methodology described in ref.~\cite{Vaitiekenas2020Feb}. In brief, in the dirty limit of superconductors, the coherence length is given by $\xi_{\mathrm{S}} = \sqrt{\pi\hbar v_{\mathrm{F}}l_e/24 k_{B}T_{c}(B = 0)}$, where $v_{\mathrm{F}} = 2.03 \times 10^{6}$ m/s is the electron Fermi velocity in Al, and $l_e$ is the mean free path. This latter parameter is obtained from the resistivity of the Al shell. By taking $R_n$, and considering the geometrical dimensions of the shell in each of our devices, we estimate $l_e \sim 2$ nm. From this value, we calculate $\xi_{\mathrm{S}}$ for the 5 measured nanowires, obtaining a distribution  in the range of 75-105 nm, consistent with the values obtained from the AG fitting of the dips in the main text. 

\subsection{Features related to the superconductivity of the Ti/Al contacts 
}

In this section, we will discuss dip-related features that are observed in all devices with superconducting lithographic contacts (Ti/Al), but that are absent when the contacts are normal (Cr/Au). 

\begin{figure}[H]
    \centering
    \includegraphics[width=16cm]{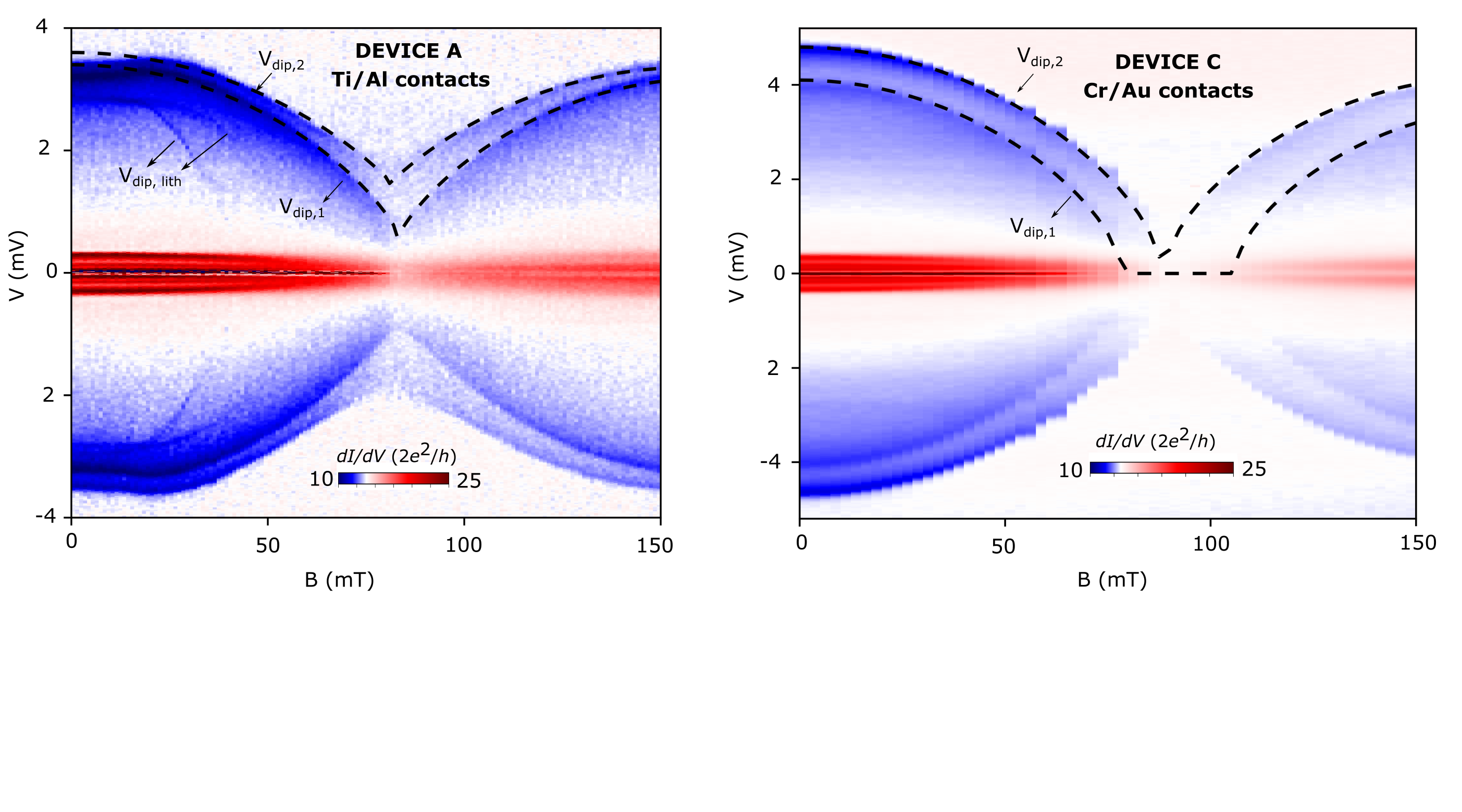}
    \caption{\textbf{Dip features in devices with superconducting (left panel) and normal (right panel) lithographic contacts}. Devices with Ti/Al contacts show additional faint dips that are suppressed for low magnetic fields. They also show a slight increase in $V_{dip,i}$ upon applying $B$ from zero to $\sim$ 20 mT.}
    \label{fig:AlvsAu}
\end{figure}

We start by addressing the
faint $dI/dV$ dips that were mentioned in passing in the discussion of Fig.~2b (labeled as $V_{dip, lith}$). These dips are more prominently seen in measurements taken as a function of $T$ or $B$. Indeed, they are also present in the $B$-field dependences in Figs.~3a and 4b, although their visibility is compromised by the lower resolution of those measurements. We show in Fig.~\ref{fig:AlvsAu}a a higher resolution $dI/dV(V, B)$ measurement for device A, focusing on lower magnetic fields. This measurement is similar to Fig.~3a, but it was taken in a different cool-down.  For this reason, we note that even though both measurements were taken at the same gate voltage ($V_g = 80$ V), $R_J$ (and consequently $V_{dip,i}$) are slightly different owing to a small shift in the pinch-off voltage of the device upon thermal cycling. Interestingly, $R_{lead,i}$ remains unchanged for the different cool-downs, reinforcing that it is a property of the leads and not of the junction. Importantly, we note that the behavior of the faint dips is consistent with the superconductivity of 240 nm-thick Al films with lateral dimensions $\sim \mu$m. Notably, their critical temperature ($T_{c, lith}(B = 0) \approx. 1.1$ K) is lower than that of the epitaxial shell ($T_{c,i}(B= 0) \approx 1.35$ K), and their critical magnetic field is $\sim 20-50$ mT. We thus conclude that the faint dips indeed have their origin in the lithographic Ti/Al contacts. We do not discuss these dips further, as they do not affect the main conclusions of this work.  

We also attribute the slight increase of $V_{dip,i}$ at low fields (up to $\sim 20$ mT) to the Ti/Al contacts. As we mentioned in the main text, this effect leads to a small discrepancy between the data and the AG fitting.
Fig.~\ref{fig:AlvsAu} clearly demonstrates that the dips in devices with Cr/Au contacts do not show such a discrepancy at low $B$.
 In analogy to the previous effect, we speculate that the present behavior is also related to the superconductor-to-normal transition of the Ti/Al film. 
In brief, we believe that the closing of the superconducting gap of the Ti/Al contacts slightly improves the thermal transport from the junction to the bath, leading to a small renormalization of $R_{lead,i}$. Indeed, we estimate that $R_{lead,i}$ at $B = 0$ is approx. 10\% higher than at $B = 20$ mT, suggesting a slightly higher thermal resistance.

\newpage

\subsection{Determining device parameters}\label{sec:Device parameters}
In this section we provide detail on the fitting of parameters for device A, B and C. From the main text it is already established how zero-field critical temperature, $T_{c}(B=0)$, and lead resistance, $R_{lead}$, are obtained by monitoring dips under change of cryostat temperature, $T_{bath}$, and junction gate, $V_g$, respectively. Additionally, for a given $V_g$ we measure the zero-field normal resistance, $R_J$, and maximal excess current, $\max(I_{exc,1}(V)+I_{exc,2})(V)$, which we use to fit the number of transmission channels, $N$, and the transmission of each channel, $\tau$, as to produce the same ratio of excess current to resistance in theory calculations. Realistically, 
each channel, $j$, will have a different transmission and fitting each $\tau_j$ can be achieved by precise fitting of MAR peaks \cite{Goffman2017Sep}. As we primarily focus on high-bias measurements, and to keep the number of fitting parameters low, we deem this procedure not worthwhile. 

Next, we elaborate on the fitting of the Little-Parks lobes observed in $V_{dip}(B)$ as a function of field, and compare it to expected wire parameters. Little-Parks oscillations of $T_{c}(B)/T_{c}(0) \approx V_{dip}(B)/V_{dip}(0)$ in a superconducting thin cylinder in the dirty limit, is described by \cite{AbrikosovGorkov1961, Skalski1964Dec},
\begin{equation}
\label{eqn:AGE}
\ln\left(\frac{T_{c}({\alpha})}{T_{c}(0)}\right) = \Psi\left(\frac{1}{2}\right)
- \Psi\left(\frac{1}{2} + \frac{\alpha}{2\pi k_{B}T_{c}(\alpha)} \right),
\end{equation}
where $\Psi$ is the digamma function, and $\alpha$ the pair-breaking parameter. 
As a perfect mechanical alignment between the nanowire axis and the applied magnetic field is not experimentally feasible we leave a small angle, $\theta$, as an additional fitting parameter, resulting in a parallel and a perpendicular contribution to the magnetic field: $B_{\|} = B\cos{\theta}$, $B_{\perp} = B\sin{\theta}$. Difference in $\theta$ between two leads we attribute to a possible curvature of the nanowire. Consequently, the total pair-breaking is given by $\alpha = \alpha_{\|} + \alpha_{\perp}$ \cite{Vekris2021Sep,Vaitiekenas2020Feb} with,
\begin{align}
\alpha_{\|}&=\frac{4 \xi_{\mathrm{S}}{ }^{2} T_{c}(0)}{A}\left[\left(n-\frac{\Phi_{\|}}{\Phi_{0}}\right)^{2}+\frac{t_{\mathrm{S}}^{2}}{d^{2}}\left(\frac{\Phi_{\|}^{2}}{\Phi_{0}^{2}}+\frac{n^{2}}{3}\right)\right], \\
\alpha_{\perp}&=\frac{4 \xi_{\mathrm{S}}{}^{2} T_{c}(0)\lambda}{A} \frac{\Phi_{\perp}^{2}}{\Phi_{0}^{2}}.
\end{align}
Here $n$ denotes the fluxoid quantum number, $\Phi_{\|} = B_{\|}A$, $\Phi_{\perp} = B_{\perp}A$, $A=\pi d^2/4$, and $\lambda$ is a free fitting parameter determining the perpendicular contribution to pair-breaking.
For the purpose of fitting this function is characterized by the following four components,
\begin{align}
B_{p} = \frac{\Phi_0}{A\cos\theta}, \hspace{0.3cm} C_1 = \frac{4\xi_S^2 T_{c}(0)}{A}, \hspace{0.3cm} C_2 = \frac{1}{3}\frac{t_S^2}{\pi A} + \lambda \frac{\sin^2\theta}{\cos^2\theta}, \hspace{0.3cm} C_3 = \lambda A^2,
\end{align}
where $B_{p}$ is the measured LP periodicity, $C_1$ sets the amplitude of periodic oscillations, $C_2$ the decay at integer flux, $\Phi_\| / \Phi_0 = n$, and $C_3$ the decay for a perpendicular field ($\theta \approx \pi/2)$. A given measurement of $V_{dip}$ as a function of parallel  magnetic field, possibly with a small $\theta$, in combination with a perpendicular field measurement with $\theta \approx \pi/2$, can be fitted by the components $\{ B_{p}, C_1, C_2, C_3\}$, and consequently any parameters yielding identical $\{ B_{p}, C_1, C_2, C_3\}$ also provides a fit. Here we assume perfect alignment in the perpendicular direction as a small parallel component is negligible, while a small perpendicular component to a parallel alignment is not. If we assume that $\{A, t_S, \xi_S, \lambda, \theta \}$ are all free parameters a unique fit cannot be obtained. Nonetheless, the space of possible fits for dips 1\&2 in device A and dip 2 in device B is restricted to $\theta \in \{ 0^\circ,10^\circ \}$ as shell thickness, $t_S$, otherwise becomes complex in order to keep $C_2$ constant.

By fixing $t_S = 15$~nm (from fabrication $t_S\approx 20$~nm) and $\lambda = 1.7$ a unique fit is obtained for all dips, with corresponding values shown in tables below. The resulting fits for all devices can be seen in Extended Data Fig.~1-2. For this choice, we find from $B_p$ that $d_{A},\hspace{0.1cm} d_C \approx 125$~nm and $d_B\approx 105$~nm (with $A$, $B$ and $C$ indicating device) comparable to the nominal length of $135$~nm from fabrication. In the allowed range of freedom for $\theta$, parameters $\{A, \xi_S, \lambda\}$ only varies within third digit precision, consequently we can conclude that the coherence length, $\xi_S$, must be different between lead 1 and 2 in order to obtain a good fit. This highlights the ability of Joule spectroscopy to extract the coherence lengths of each lead independently. Device B lead 2 is a special case as no Little-Parks oscillation is observed, and we concluded that the Al shell is not doubly connected. As a function of $B_\|$ a monotone decaying trend of $V_{dip,2}$ is observed which is fitted by setting $\alpha_\| = 0$ and fitting $\theta$. Consequently, the angle, $\theta$, for dip 2 in device B should only be understood as a fitting parameter since we lack knowledge of the state of the Al shell.

\newpage
\begin{table}[ht]
	\begin{center}
		\begin{tabularx}{460pt}{@{} |c *{10}{>{\Centering}X}| @{}}
                \multicolumn{11}{c}{\textbf{Device A}} \\
			\hline
			\text{lead} & $T_{bath}\left[\text{K}\right]$ & $N$ & $\tau$ & $R_{lead}\left[\Omega\right]$ &  $T_{c}(0)\left[\text{K}\right]$ & $\xi_{S}\left[\text{nm}\right]$ & $B_{p}\left[\text{mT}\right]$ & $\lambda$ & $t_{S} \left[\text{nm}\right]$ & $\theta\left[\text{deg}\right]$ \\ \hline
			
			\text{1} & $0.25$ & $16$ & $0.675$ & $4.4$ &  $1.35(1.4)$ & $100$ & $166$ & $1.7$ & $15$ & $3.7$ \\
			
			\text{2} & $0.25$ & $16$ & $0.675$ & $3.8$ &  $1.35(1.4)$ & $90$ & $162$ & $1.7$  & $15$ & $5.6$ \\ \hline
		\end{tabularx}
		\caption{\textbf{Parameters of Device A}. $T_{bath}$, $N$ and $\tau$ are all tuneable, 
            and values shown here correspond to those in Fig.~3. The $T_{c}(0)$ value in parentheses is the one used in theory calculations. Other quantities are given by: $\Delta_{i}(0) = 1.76 k_B T_{c,i}(0)$, $R_J = 1/G_0N\tau$, and $A = \Phi_0/B_p\cos\theta$.}
		\label{tab:DevA}
	\end{center}

 	\begin{center}
		\begin{tabularx}{460pt}{@{} |c *{10}{>{\Centering}X}| @{}}
                \multicolumn{11}{c}{\textbf{Device B}} \\
			\hline
			\text{lead} & $T_{bath}\left[\text{K}\right]$ & $N$ & $\tau$ & $R_{lead}\left[\Omega\right]$ &  $T_{c}(0)\left[\text{K}\right]$ & $\xi_{S}\left[\text{nm}\right]$ & $B_{p}\left[\text{mT}\right]$ & $\lambda$ & $t_{S} \left[\text{nm}\right]$ & $\theta\left[\text{deg}\right]$ \\ \hline
			
			\text{1} & $0.01$ & $15$ & $0.69$ & $2.0$ & $1.35(1.4)$ & $75$ & $225$ & $1.7$ & $15$ & $7$ \\
			
			\text{2} & $0.01$ & $15$ & $0.69$ & $0.7$ & $1.35(1.4)$ & $65$ & $225$ & $1.7$ & $15$ & $11$ \\ \hline
		\end{tabularx}
		\caption{\textbf{Parameters of Device B}. $T_{bath}$, $N$ and $\tau$ are all tuneable, 
            and values shown here correspond to those in Fig.~4. The $T_{c}(0)$ value in parentheses is the one used in theory calculations. Other quantities are given by: $\Delta_{i}(0) = 1.76 k_B T_{c,i}(0)$, $R_J = 1/G_0N\tau$, and $A = \Phi_0/B_p\cos\theta$. Note that for lead 2 we put $\alpha_\|(B) = 0$ and $B_p$ is fitted to yield the correct perpendicular decay for $\lambda = 1.7$. The angle, $\theta$, should only be regarded as a fitting parameter for lead 2.
            }
		\label{tab:DevB}
	\end{center}

 	\begin{center}
		\begin{tabularx}{460pt}{@{} |c *{10}{>{\Centering}X}| @{}}
                \multicolumn{11}{c}{\textbf{Device C}} \\
			\hline
			\text{lead} & $T_{bath}\left[\text{K}\right]$ & $N$ & $\tau$ & $R_{lead}\left[\Omega\right]$ &  $T_{c}(0)\left[\text{K}\right]$ & $\xi_{S}\left[\text{nm}\right]$ & $B_{p}\left[\text{mT}\right]$ & $\lambda$ & $t_{S} \left[\text{nm}\right]$ & $\theta\left[\text{deg}\right]$ \\ \hline
			
			\text{1} & $0.25$ & $21$ & $0.64$ & $1.8$ &  $1.0(1.0)$ & $115$ & $182$ & $1.7$ & $15$ & $5.8$ \\
			
			\text{2} & $0.25$ & $21$ & $0.64$ & $2.4(2.7)$ &  $1.33(1.4)$ & $100$ & $176$ & $1.7$  & $15$ & $7.7$ \\ \hline
		\end{tabularx}
		\caption{\textbf{Parameters of Device C}. $T_{bath}$, $N$ and $\tau$ are all tuneable, 
            and values shown here correspond to those in Fig.~4. The $T_{c}(0)$ and $R_{lead}$ value in parentheses is the one used in theory calculations. Difference in $R_{lead}$ stems from difference in $T_{c}(0)$. Other quantities are given by: $\Delta_{i}(0) = 1.76 k_B T_{c,i}(0)$, $R_J = 1/G_0N\tau$, and $A = \Phi_0/B_p\cos\theta$.}
		\label{tab:DevC}
	\end{center}
\end{table}

\newpage

\subsection{Cooling power by electron-phonon coupling}

We estimate here the cooling power provided by electron-phonon coupling in the epitaxial Al shell, $P_{\textit{e-ph}}$, to support our assumption that, in our devices, cooling predominantly occurs via quasiparticles in the leads. Following refs.~ \cite{Wellstood94,Courtois2008}, we write the heat balance equation:

\begin{equation}
    \label{eqn:elec-ph}
    P_{\textit{e-ph}} = \Sigma U(T_{el}^{5} - T_{ph}^{5}),
\end{equation}

 where $\Sigma = 1.8$ nW/$\mu$m$^3$ K$^5$ is the Al electron-phonon coupling parameter \cite{Maisi2013}, $U \approx 7.07 \times 10^{-3} \mu$m$^3$ is the volume of the Al shell (assuming a NW core diameter of 135 nm, a shell thickness of 15 nm, and a length of 1 $\mu$m), $T_{el}$ is the electron temperature, and $T_{ph}$ is the phonon temperature, which we take to be equal to $T_{bath}$. At the superconductor-to-normal metal transition of the leads, the electron temperature reaches the superconducting critical temperature, $T_c = 1.35$ K. By assuming $T_{ph} = 0.25$ K, we obtain $P_{\textit{e-ph}} \sim$ 0.057 nW, which is more than two orders of magnitude lower than the measured $P_{dip,i} \sim$ 10 nW. We therefore conclude that heat diffusion by quasiparticles in the leads is a more efficient cooling mechanism in our devices.

\newpage

\section{Transport theory}
In this section we elaborate on the main theoretical results relating the measurements of high voltage conductance dips with properties of the junction and leads.Simple approximate relations connecting the conductance dips with lead and junction parameters, such as $T_c$, are derived by assuming that thermal transport is solely mediated by lead quasi-particles, and that for a given power input each lead independently reaches thermal equilibrium. Finally to validate these relations we self-consistently calculate the power each lead receives from joule heating through the use of Keldysh-Floquet transport methodology, accounting for both pair-breaking, asymmetric leads, and Andreev reflection to all orders. Results from this approach are compared to experimental data both in the supplement and in the main text.    

\subsection{Pair-broken superconductor}
The application of either a parallel or perpendicular magnetic field induces a pair-breaking, $\alpha$, in the leads, and because of the small mean-free path compared to coherence length the resulting pair-broken superconductivity can be described by Abrikosov-Gor'kov theory \cite{AbrikosovGorkov1961,Skalski1964Dec,Larkin1965}. In this subsection we iterate the key components of this theory used in our calculations. Under the influence of pair-breaking, the quasi-classical retarded Green function is given by
\begin{equation}
g^R(\omega) = -i\pi\nu_F \frac{u(\omega) - \tau_x}{\sqrt{u(\omega)^2 - 1}}, \label{PBSC:GR}
\end{equation}
where $\nu_F$ is the density of state at the Fermi level and $\tau_x$ a pauli matrix in Nambu space. The complex number $u(\omega)$ is obtained as the solution of
\begin{equation}
u(\omega)\Delta(\alpha, T) = \omega + i\alpha\frac{u(\omega)}{\sqrt(u(\omega)^2 - 1)}. \label{PBSC:equ}
\end{equation}
For a given $\Delta(\alpha,T)$ this equation can be expressed as a fourth order polynomial with root $u(\omega)$ chosen as to satisfy appropriate boundary conditions of the Green function. For the pairing parameter self-consistency with the Green function demands,
\begin{equation}
\Delta(\alpha, T) = \nu_F U\int_{0}^{\hbar\omega_D} d\omega\hspace{0.1cm} \text{Re}\frac{1}{\sqrt{u(\omega)^2-1}}\tanh{\frac{1}{2}\frac{\omega}{k_B T}}, \label{PBSC:IdealeqDelta}
\end{equation}
where $U$ is the strength of the interaction, assumed weak, $T$ denotes temperature, and $\omega_D$ the Debye frequency. The various scales appearing in this problem are connected by standard BCS relations; $\Delta_0 = 2\hbar\omega_De^{-1/\nu_FU}$ and $k_B T_{c0} = \frac{2e^\gamma}{\pi}\hbar\omega_D e^{-1/ \nu_F U}$ with $T_{c0} = T_c(\alpha=0)$, $\Delta_0 = \Delta(\alpha=0, T=0)$ and $\gamma$ denoting Euler's constant. For finite pair-breaking and zero-temperature a closed form solution of $\Delta(\alpha, 0)$ exist, 
\begin{equation}
\ln{\frac{\Delta_0}{\Delta(\alpha,0)}} = 
\begin{cases}
-\frac{\pi}{4}\frac{\alpha}{\Delta(\alpha,0)} & \text{if }\alpha \leq \Delta(\alpha,0), \\
-\ln{\frac{\alpha+\sqrt{\alpha^2-\Delta(\alpha,0)^2}}{\Delta(\alpha,0)}} + \frac{\sqrt{\alpha^2-\Delta(\alpha,0)^2}}{2\alpha} -\frac{1}{2}\arctan{\frac{\Delta(\alpha,0)}{\sqrt{\alpha^2-\Delta(\alpha,0)^2}}} & \text{if }\alpha\geq\Delta(\alpha,0),
\end{cases}
\end{equation}
\begin{figure}[H]
    \centering
    \includegraphics[width=16cm]{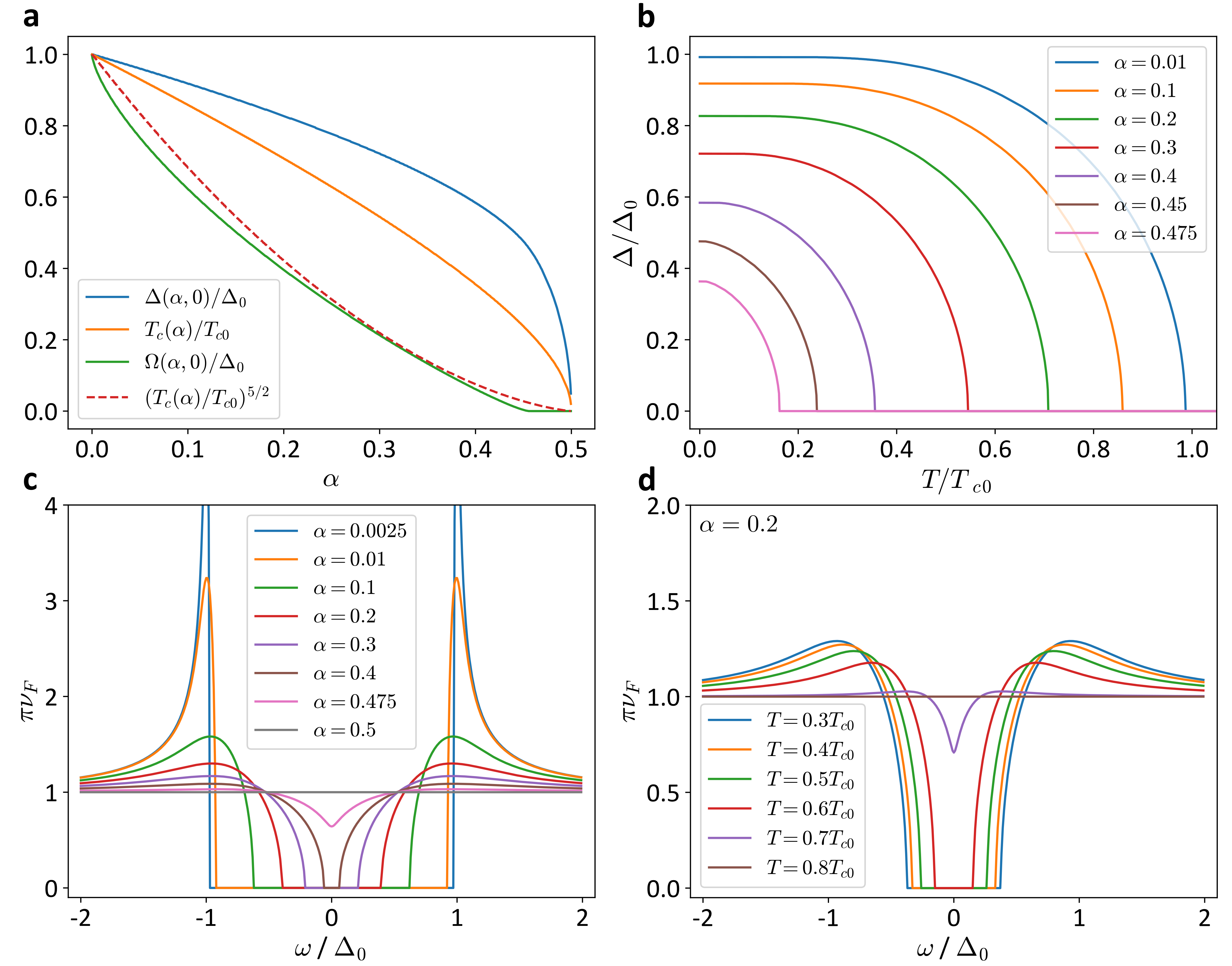}
    \caption{\textbf{Effects of pair-breaking}. \textbf{a} Pairing parameter $\Delta(\alpha, 0)$, critical temperature $T_c(\alpha)$, and the spectral gap $\Omega(\alpha,0)$ as a function of pairing. \textbf{b} Numerical solutions of eq.~(\ref{PBSC:eqDelta}) for $\Delta(\alpha,T)$ for various $\alpha$. \textbf{c} Spectral function $A(\omega)=-\text{Im}\hspace{0.1cm}g^R_{11}(\omega)$ at zero temperature. \textbf{d} The effect of temperature on the spectral function for finite pair-breaking. In all plots $\alpha$ is in units of $\Delta_0$.}
    \label{fig:Pair-Breaking}
\end{figure}
which can be solved by intersect. In the case of finite temperature eq.~(\ref{PBSC:IdealeqDelta}) has to be solved as an integral equation, and using BCS relations we express it as, 
\begin{equation}
\frac{2\Delta(\alpha,T)}{\Delta_0}\log{N}= \int_{0}^{N} dx\hspace{0.1cm}\text{Re} \frac{1}{\sqrt{u(\Delta_0 x/2)^2-1}} \tanh{\frac{\Delta_0 x}{4T}}, \label{PBSC:eqDelta}
\end{equation}
where $N$ is numerical parameter chosen sufficiently large number as to assure the integrand approaches $\frac{2\Delta(\alpha,T)}{\Delta_0 x}$ for $x\rightarrow N$. For a given $\alpha$ and $T$ eq.~(\ref{PBSC:equ}) and eq.~(\ref{PBSC:eqDelta}) can be jointly solved numerically to obtain $\Delta(\alpha,T)$ and $u(\omega)$, with the size of $N$ determining precision.
The above relations allow evaluation of the retarded Green function, eq.~(\ref{PBSC:GR}), for any value of $\alpha$ and $T$ from which the spectral function $A(\omega) = -\text{Im}\hspace{0.1cm}g^R_{11}(\omega)$ can be obtained. One characteristic of a pair-broken supercondcutor is that the spectral gap, denoted $\Omega(\alpha,T)$, is not equal to the pairing parameter, $\Delta(\alpha, T)$, as in the case of BCS superconductivity but instead given by,
\begin{equation}
\Omega(\alpha,T) = \left(\Delta(\alpha, T)^{\frac{2}{3}} - \alpha^{\frac{2}{3}} \right)^{\frac{3}{2}}.
\end{equation}
In Fig.~\ref{fig:Pair-Breaking} we show various quantities characterizing superconductivity dependence on pair-breaking and temperature. In Fig.~\ref{fig:Pair-Breaking}a an approximate relation relating spectral gap to critical temperature, $\Omega(\alpha,0)/\Delta_0 \approx \left(T_c(\alpha)/T_{c0}\right)^{5/2}$, is additionally shown.  

\subsection{Lead thermal balance}
As a consequence of electron tunneling across the junction a non-equilibrium distribution of high energy quasi-particles emerge on the left and right lead. In the following we assume that on a given lead this distribution relaxes to an equilibrium distribution releasing a power $P$ at the lead interface. We further assume that all heat diffusion through the epitaxial aluminium stems from activated quasi-particles and solve for thermal equilibrium. This derivation largely follows calculations of \textit{Tomi et al.} \cite{Tomi2021Oct}, here expanded to also include pair-breaking.

We model the epitaxial aluminium leads as a 1D wire of length $L$ and cross sectional area $S$. Thermal equilibrium requires that the power passing through each segment of wire be equal, such that a lead temperature distribution, $T(x)$, stabilizes. This condition amounts to the heat diffusion equation,
\begin{equation}
S\kappa_S(\alpha,T)\frac{dT}{dx} = -P, \label{Tomi:Difeq}
\end{equation}
with the thermal conductivity, $\kappa_S(\alpha,T)$, given by the analogous Wiedemann-Franz law for a pair-broken superconductor \cite{Ambegaokar1965Feb},
\begin{equation}
\kappa_S(\alpha,T) = \frac{4k_B^2\sigma}{e^2}T\int_{\frac{\Omega(\alpha,T)}{2k_BT}}^\infty dx\hspace{0.1cm} \frac{x^2}{\cosh^2{x}}h(2k_BT x ,\alpha,T)
\end{equation}
where the effect of pair-breaking is encapsulated in the function,

\begin{figure}[H]
    \centering
    \includegraphics[width=16cm]{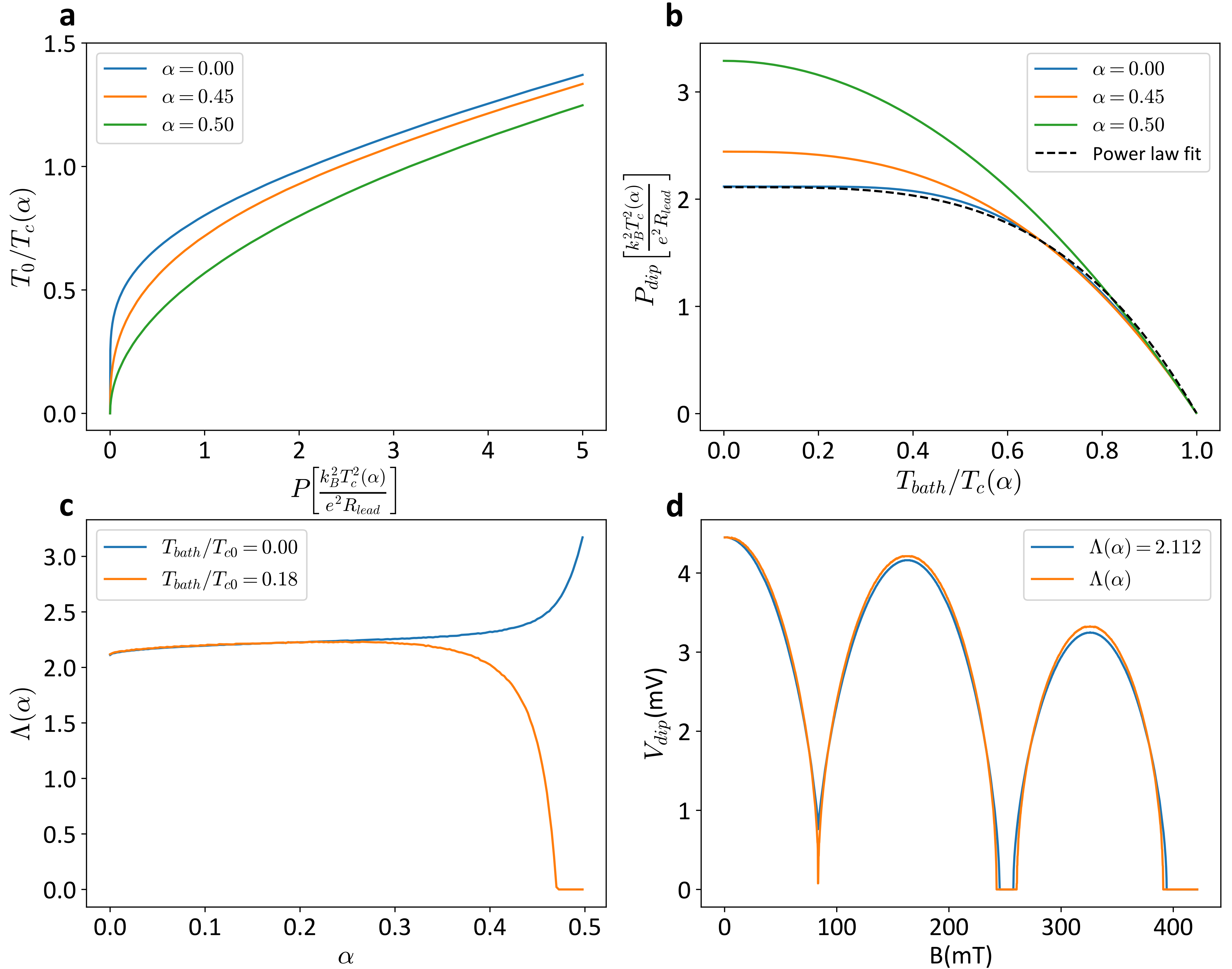}
    \caption{\textbf{Solutions to quasi-particle heat diffusion}. \textbf{a} Interface temperature as a function of injected power from eq.~(\ref{eq:PH-PtoT}). \textbf{b} Scaling of dip power, $P_{dip}$, with bath temperature, $T_{bath}$, obtained from eq.~(\ref{eq:PH-Pdip}). \textbf{c} $\Lambda(\alpha, T_{bath})$ as a function of pair-breaking for zero temperature and $T_{bath} = 0.18 T_{c0}$ corresponding to $T_{bath} = 250$~mK and $T_{c0} = 1.4$~K. \textbf{d} Expected position of $V_{dip,1}$ (obtained from eq.~(\ref{eq:Vdip}) using device A parameters, see table~\ref{tab:DevA}) calculated with $\Lambda(\alpha, T_{bath})$ assumed constant and by using eq.~(\ref{eq:PH-Lambda}). $\alpha$ is in units of $\Delta_0$.}
    \label{fig:PHeatBalance}
\end{figure}

\begin{equation}
h(\omega, \alpha, T) = \left(\text{Re}\frac{u(\omega)}{\sqrt{u(\omega)^2-1}}\right)^2-  \left(\text{Re}\frac{1}{\sqrt{u(\omega)^2-1}}\right)^2,
\end{equation}
$h(\omega, \alpha, T)$ depends on $\alpha$ and $T$ through $u(\omega)$'s dependence of $\Delta(\alpha,T)$ in eq.~(\ref{PBSC:equ}). Integrating eq.~(\ref{Tomi:Difeq}) across the length of the wire and imposing the boundary condition $T(x=L) = T_{bath}$ and $T(x=0) = T_0$, with $T_{bath}$ denoting environment temperature and $T_0$ temperature at the junction interface, yields,
\begin{equation}
P = \frac{8k_B^2}{e^2 R_{lead}}\int_{T_{bath}}^{T_0} dT\hspace{0.1cm}T \int_{\frac{\Omega(\alpha,T)}{2k_BT}}^\infty dx\hspace{0.1cm} \frac{x^2}{\cosh^2{x}}h(2k_BTx,\alpha,T), \label{eq:PH-PtoT}
\end{equation}
with lead resistance defined as $R_{lead}=2L/\sigma S$. A conductance dip occurs whenever $T_0 = T_c(\alpha)$ and the required power can be expressed as,
\begin{equation}
P_{dip} = \Lambda(\alpha, T_{bath})\frac{k_B^2T^2_c(\alpha)}{e^2 R_{lead}} \label{eq:PH-Pdip}
\end{equation}
with the thermal properties of the leads described by the unitless function,
\begin{equation}
\Lambda(\alpha, T_{bath}) = \frac{8}{T_c^2(\alpha)}\int_{T_{bath}}^{T_c(\alpha)} dT\hspace{0.1cm}T \int_{\frac{\Omega(\alpha,T)}{2k_BT}}^\infty dx\hspace{0.1cm} \frac{x^2}{\cosh^2{x}}h(2k_BTx,\alpha,T). \label{eq:PH-Lambda}
\end{equation}

This function is bounded by $\pi^2/3\geq\Lambda(\alpha,T)\geq 0.0$, with the lower bound reached for $T_{bath}\geq T_c(\alpha)$ when no additional power is required to drive the interface normal, and the upper bound reached for $\alpha/\Delta_0 = 0.5$ when the lead becomes metallic and most thermally conductive. In the zero temperature BCS limit $\Lambda(0, 0) = 2.112$ \cite{Tomi2021Oct} and remains approximately constant so far $T_c(\alpha) \geq T_{bath}$, as shown in Fig~\ref{fig:PHeatBalance}c-d. For $\alpha = 0$ one obtains the approximate power law,
\begin{equation}
\Lambda(0, T_{bath}) \approx 2.112\left(1 - \frac{T_{bath}^{3.6}}{T_{c0}^{3.6}}\right),
\end{equation}
which is compared to the exact curve in Fig.~\ref{fig:PHeatBalance}b. The fitted power, $3.6$, attempts to bridge the transition from an initial exponentially suppressed curve for $T_{bath}\ll T_{c0}$ to a second order closing, $T_{bath}^2/T_{c0}^2$, at $T_{bath}\approx T_{c0}$ \cite{Tomi2021Oct}.

\subsection{Schematic Theory for dips}
Next, we present a schematic calculation to obtain the bias position of the dips. In the high bias regime, $eV \gg \Delta_1 + \Delta_2$ with $\Delta_i = \Delta_i(\alpha_i = 0, T_{bath}=0)$ for lead 1 and 2, the excess current can be described as originating from two independent S - N junctions with the total current across the junction given by,
\begin{equation}
I = \frac{V}{R_J} + I_{exs,1}(\Delta_1 ,\alpha_{1}, T_{0,1}) + I_{exs,2}(\Delta_2 ,\alpha_2, T_{0,2}),
\end{equation}
where excess current, $I_{exs,i}$, depends non-trivially on both temperature and pair-breaking. The power deposited on either lead is given by,
\begin{equation}
P_{1(2)} = \frac{V^2}{2R_J} + V I_{exs,2(1)}(\Delta_{2(1)},\alpha_{2(1)},T_{0,2(1)}).
\end{equation}
To obtain interface temperature $T_{0,i}$ exactly requires a self-consistent treatment; for a given $P_{i}$ one finds $T_{0,i}$ from eq.~(\ref{eq:PH-PtoT}), but a change of $T_{0,i}$ modifies $P_{i}$. If the normal contribution to current greatly exceeds the excess current at a thermal dip, $V_{dip,i}/R_J \gg I_{exs,1},\hspace{0.1cm} I_{exs,2}$, this self-consistency is negligible as $P_{1} \approx P_{2} \approx  V_{dip,i}^2/2R_J$ yielding,
\begin{equation}
V_{dip,i} = R_J I_{dip,i} = \sqrt{2\Lambda(\alpha_i,T_{bath})}\sqrt{\frac{R_J}{R_{lead,i}}}\frac{k_B T_{c,i}(\alpha_i)}{e}, \label{eq:Vdip}
\end{equation}
identical to eq.~(3) of the main text. Under the application of a magnetic field both $\Lambda(\alpha_i,T_{bath})$ and $T_{c,i}(\alpha)$ are simultaneously modified, but as $\Lambda(\alpha_i, T_{bath})$ can be approximated as a constant (see Fig.~\ref{fig:PHeatBalance}d) changes of $V_{dip,i}$ directly correspond to changes of $T_{c,i}$. These equations constitute the main results enabling Joule spectroscopy.

\subsection{Keldysh-Floquet transport theory}
In this subsection we use the Keldysh-Floquet Green function technique for a pair-broken superconductor \cite{Zaitsev1998Apr} to self-consistently in $T_{0,i}$ calculate DC current, $I$, plotted in theory figures of the main text. These calculations additionally support that previous assumptions of constant $\Lambda(\alpha_i, T_{bath})$ and $V_{dip,i}/R_J \gg I_{exs,i}$ is reasonable, and allow us to compare low-bias MAR structure with high bias dips. We consider transport to occur between the left and right Al superconducting shell, which are described by quasi-classical Green functions, and model the junction as a generic contact with $N$ transmission eigenvalues $\tau_i$ of the corresponding normal-state scattering matrix. Using appropriate boundary conditions for the quasi-classical Greens functions, transport can be described via the matrix current, \cite{Nazarov1999May, Virtanen2014Jul},
\begin{equation}
\check{I}(t) = \frac{e^2}{h}\sum_n\frac{\tau_n \left[\check{g}_1,\check{g}_2\right]_-}{1-\frac{1}{2}\tau_n +\frac{1}{4}\tau_n \left[\check{g}_1,\check{g}_2\right]_+}\left(t,t\right)
\end{equation}
where $-(+)$ describe (anti-)commutators and with time-convolution assumed in the matrix structure, $\check{g}_1\check{g}_2(t,t') = \int_{-\infty}^{\infty}dt''\check{g}_1(t,t'')\check{g}_2(t'',t')$. The Green functions are written in Nambu-Keldysh space,
\begin{equation}
\check{g}_i = \begin{pmatrix}
\bar{g}_i^R && \bar{g}_i^K \\
0 && \bar{g}_i^A 
\end{pmatrix},\hspace{0.4cm} \bar{g}_1(t,t') = \frac{\tau_z}{i\pi \nu_{F,1}} g_1(t-t'), \hspace{0.4cm} \bar{g}_2(t,t') = \frac{\tau_z}{i\pi\nu_{F,2}}e^{ieVt\tau_z/\hbar}g_2(t-t')e^{-ieVt'\tau_z/\hbar} 
\end{equation}
where $g^R_i(t-t') = \int_{-\infty}^\infty d\omega\hspace{0.1cm} g^R_i(\omega)e^{-i\omega(t-t')}$ and $g^R_i(\omega)$ is given by eq.~(\ref{PBSC:GR}) with $g^A_i(\omega) = \left[g^R_i(\omega)\right]^\dagger$ and $g^K_i(\omega) = \left(g^R_i(\omega)-g^A_i(\omega)\right)\tanh\left(\omega/2T_{0,i}\right)$. In this framework the Green functions of lead $i$ are completely specified by parameters $\{\Delta_{i}, \nu_{F,i}, \alpha_i, T_{0,i}\}$.  The gauge part, $e^{ieVt\tau_z/\hbar}$, originates from the AC Josephson effect where an applied DC voltage drop creates explicit time-dependence, and where $\tau_z$ denotes a pauli matrix in Nambu space. To highlight the connection between the quasi-classical and tunneling descriptions we rewrite the matrix current using Dyson series,
\begin{equation}
\check{I}(t) = \frac{4e^2}{h}\sum_n\frac{\tau_n}{4-2\tau_n} \left[\check{g}_1,\check{g}_2\right]_- \check{M}_{+n} = \frac{4e^2}{h}\sum_n b_n\left(\check{g}_2\check{g}_1 \check{M}_{21,n} - \check{g}_1\check{g}_2 \check{M}_{12,n}\right)
\end{equation}
with $\tau_n = 4b_n^2/(1+b_n)^2$ and,
\begin{equation}
\check{M}_{+n} = 1-\frac{\tau_n}{4-2\tau_n}\left[\check{g}_1,\check{g}_2\right]_+ \check{M}_{+n}, \hspace{0.3cm} \text{and} \hspace{0.3cm} \check{M}_{ij,n} = 1 + b_n\check{g}_i\check{g}_j \check{M}_{ij,n}.
\end{equation}
These expression are obtained by utilizing the following identity $\check{g}_i \check{g}_i = I$ and we recognize $b_n = \pi^2\nu_{F,1} \nu_{F,2} |t_n|^2 $ where $t_n$ describes the tunneling amplitude in a corresponding tunneling model. Lastly we identify $b_n\check{g}_2\check{g}_1 \check{M}_{21,n} = \sqrt{b_n} \check{G}_{21,n}$ with the dressed Green functions defined via typical equation-of-motion structure,
\begin{equation}
\check{G}_{21,n} = \check{g}_2\sqrt{b_n}\check{G}_{11,n} \hspace{0.3cm} \text{and} \hspace{0.3cm} \check{G}_{11,n} = \check{g}_1 + \check{g}_1\sqrt{b_n}\check{g}_2\sqrt{b_n}\check{G}_{11,n},
\end{equation}
such that the matrix current is given by,
\begin{equation}
\check{I}(t) = \frac{4e^2}{h}\sum_n\left(\sqrt{b_n} \check{G}_{21}(t,t) - \sqrt{b_n} \check{G}_{12}(t,t)\right),
\end{equation}
identical to equations obtained from S - S tunneling models \cite{Cuevas1996Sep}. From the matrix current we obtain the charge and energy current \cite{Virtanen2014Jul},
\begin{align}
I(t) &= \frac{1}{8}\text{Tr}\hspace{0.1cm}\tau_z \check{I}^K(t), \nonumber \\
P_L(t) &= \frac{1}{16} \text{Tr}\left[\epsilon \check{I}^K(t) + \check{I}^K(t)\epsilon \right], \\
P_R(t) &= I(t) - P_L(t), \nonumber
\end{align}
with $\check{I}^K(t)$ indicating the Keldysh component of the matrix current and $\epsilon(t,t') = i\partial_t \delta(t-t')$. Assuming that the system reach a time-periodic non-equilibrium steady state, $\check{g}_i(t,t') = \check{g}_i(t+T,t'+T)$ with $T=2\pi\hbar/eV$, we can transform time-convolutions into Floquet matrix structure. Considering only the DC component, corresponding to the zeroth Floquet band, we obtain the following equations for the currents,
\begin{align}
I &= \frac{2e^2}{\hbar}\sum_{n}\int_{-\infty}^{\infty}d\omega\text{Re}\hspace{0.1cm}\text{Tr}\left[\tau_z\hspace{0.1cm} b_n \underline{\check{M}}_{21,n}^R\left(\underline{\check{g}}_2^R\underline{\check{g}}_1^<+\underline{\check{g}}_2^<\underline{\check{g}}_1^A\right)\underline{\check{M}}_{21,n}^A\right]_{00}, \nonumber\\
P_1 &= \frac{2e^2}{\hbar}\sum_{n}\int_{-\infty}^{\infty}d\omega\text{Re}\hspace{0.1cm}\text{Tr}\left[\omega\hspace{0.1cm} b_n \underline{\check{M}}_{21,n}^R\left(\underline{\bar{g}}_2^R\underline{\bar{g}}_1^<+\underline{\bar{g}}_2^<\underline{\bar{g}}_1^A\right)\underline{\check{M}}_{21,n}^A \right]_{00}, \label{eq:KFIandP} \\
P_2 &= I - P_1, \nonumber
\end{align}
with the 'underline' indicating Floquet matrix structure and $00$ indicating initial and final Floquet index. Entrances in Floquet matrices are given by,
\begin{equation}
\bar{g}_{i,nm}^X(\omega) = \frac{1}{T}\int_{0}^T dt \hspace{0.1cm} e^{i(n-m)eVt/\hbar}\int_{-\infty}^\infty dt'\hspace{0.1cm} e^{i(\omega+meV/\hbar)(t-t')}\bar{g}_{i}^X(t,t'),
\end{equation}
with $X\in\{R,L,<\}$ and $\bar{g}^<_{i,nm}(\omega) = \left(\bar{g}^A_{i,nm}(\omega)-\bar{g}^R_{i,nm}(\omega)\right)n_F(\omega + meV/\hbar,T_{0,i})$. In this framework the product $\underline{\bar{g}}_1\underline{\bar{g}}_2$ forms a block tridiagonal matrix in Nambu-Floquet space, which $\underline{\check{M}}_{21,n}$ is a convergent series of. Consesquently for a given $b_n$ the number of included Floquet bands can be truncated to obtain $I$ and $P_i$ to any given precision. The numerical results presented in the main paper are obtained in the following way; For a given magnetic field we obtain $\alpha_{i}$ from Little-Park theory which together with an initial guess of $T_{0,i}$ yields $\Delta_i(\alpha_i, T_{0,i})$ and $g^X_i(t,t)$ via eq.~(\ref{PBSC:eqDelta}) and eq.~(\ref{PBSC:GR}). For a given bias, $eV$, we then calculate $I$ and $P_i$ using eq.~(\ref{eq:KFIandP}) including sufficient Floquent bands as to assure convergence. From $P_i$ we update $T_{i,0}$ using eq.~(\ref{eq:PH-PtoT}), which is used to update $\Delta_i(\alpha,T_{0,i})$ and $g^X_i(t,t')$ and recalculate $I$ and $P_i$ until convergence of $T_{0,i}$ is achieved. This procedure assures that thermal transport across the junction stemming from asymmetry in leads and heat diffusion is properly accounted for in a self-consistent manner.  

\begin{figure}[H]
    \centering
    \includegraphics[width=16cm]{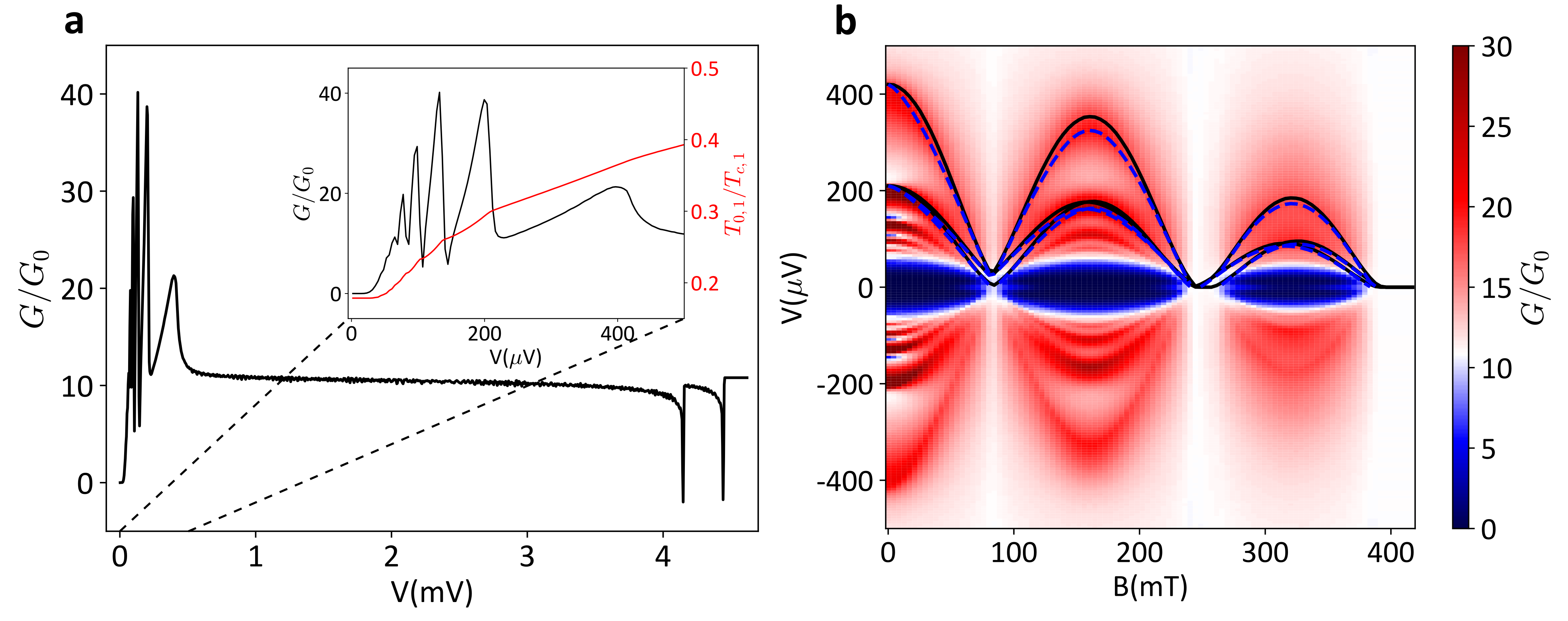}
    \caption{\textbf{Self-consistent calculation of transport}. \textbf{a} High resolution conductance line-cut at zero magnetic field for device A. Inset shows low-bias MAR structure and $T_{0,1}$ obtained self-consistently. \textbf{b} Low-bias conductance map showing the effect of magnetic field on MAR structure. Black lines and dashed blue lines indicate expected position of MAR resonances obtained from $\Omega_i(\alpha_i, 0)$ and eq.~(\ref{eq:MARlinesfromTc}) respectively. Plots are made using device A parameters, see table~\ref{tab:DevA}.
    }
    \label{fig:FloqKel}
\end{figure}

Results of self-consistent Floquet Keldysh calculations are shown both in the supplement and main text, and by using experimentally extracted parameters (see tables in subsection~\ref{sec:Device parameters}) we find a good agreement between experiment and theory. A full comparison for all devices can seen in Extended Data Figure 1. Simulations shown in both Extended Data and in the main text are performed with a finite coarse-graining set to approximately match experimental resolution. Using finer graining we find that both low-bias MAR features and high-bias conductance dips contain narrow peaks not fully resolved in experiment, as shown in Fig.~\ref{fig:FloqKel}a which is identical to Fig.~1d of the main text except graining. For a BCS superconductor with no pair-breaking MAR steps for odd $n$ appear at bias $V = (\Delta_1 +\Delta_2)/en$, and for even $n$ at $V = \Delta_i/en$. For finite pair-breaking, $\alpha_i \neq 0$, we find that MAR steps instead appear as fractions of the spectral gap, $\Omega_i(\alpha_i,T_{0,i})$, in a similar manner. In experiment, however, spectral gaps are not directly extractable from measurements of high bias dips, but as shown in Fig.~\ref{fig:PHeatBalance}a for zero temperature one approximately finds $\frac{\Omega_i(\alpha_i, 0)}{\Delta_i}\approx \left(\frac{T_{c,i}(\alpha_i)}{T_{c,i}(0)}\right)^{5/2}$ yielding MAR steps at,
\begin{equation}
V = \begin{cases}
    \frac{\Delta_1}{en}\left(\frac{T_{c,1}(\alpha_1)}{T_{c,1}(0)}\right)^{5/2} + \frac{\Delta_2}{en}\left(\frac{T_{c,2}(\alpha_2)}{T_{c,2}(0)}\right)^{5/2} & \text{if } n \text{ is odd}, \\
    \frac{\Delta_i}{en}\left(\frac{T_{c,i}(\alpha_i)}{T_{c,i}(0)}\right)^{5/2}  & \text{if } n \text{ is even}.
\end{cases} \label{eq:MARlinesfromTc}
\end{equation}
Approximating $\Lambda(\alpha, T_{bath})$ as constant renders $T_{c,i}(\alpha_i)$ proportional to $V_{dip,i}$ and consequently $T_{c,i}(\alpha)/T_{c,i}(0) = V_{dip,i}(\alpha)/V_{dip,i}(0)$. This last relation allows one to fit low-bias MAR structure directly from measurements of high-bias conductance dips. In Fig.~\ref{fig:FloqKel}b we show a simulation of low-bias conductance for device A alongside fits of MAR lines yielding good agreement between conductance peaks and MAR integers.

Lastly, it should be noted that the above analysis does not account for the low bias supercurrent branch around $V \approx 0$. Consequently the zero-bias conductance peak observed in experiment is not reproduced by numerical simulations.

\bibliography{BibSup}